**Doping-induced superconductivity in the van der Waals superatomic crystal $Re_6Se_8Cl_2$**

Evan J. Telford[1Δ], Jake C. Russell[2Δ], Joshua R. Swann[1], Brandon Fowler[2], Xiaoman Wang[3], Kihong Lee[2], Amirali Zangiabadi[4], Kenji Watanabe[5], Takashi Taniguchi[5], Colin Nuckolls[2], Patrick Batail[2,6], Xiaoyang Zhu[2], Jonathan A. Malen[3], Cory R. Dean[1]*, Xavier Roy[2]*

1 Department of Physics, Columbia University, New York, NY 10027 USA

2 Department of Chemistry, Columbia University, New York, NY 10027 USA

3 Department of Mechanical Engineering, Carnegie Mellon University, Pittsburgh, PA 15213, USA

4 Department of Applied Physics and Applied Mathematics, Columbia University, New York, NY 10027, USA

5 National Institute for Materials Science, 1-1 Namiki, Tsukuba, 305-0044 Japan

6 Laboratoire MOLTECH, CNRS UMR 6200, Université d'Angers, 49045 Angers, France

Δ Equally contributing authors

* Corresponding authors: xr2114@columbia.edu, cd2478@columbia.edu

**Superatomic crystals are composed of discrete modular clusters that emulate the role of atoms in traditional atomic solids[1–4]. Owing to their unique hierarchical structures, these materials are promising candidates to host exotic phenomena, such as superconductivity and magnetism that can be revealed through doping[5–10]. Low-dimensional superatomic crystals hold great promise as electronic components[11,12], enabling these properties to be applied to nanocircuits, but the impact of doping in such compounds remains unexplored. Here we report the electrical transport properties of $Re_6Se_8Cl_2$, a two-dimensional superatomic semiconductor[13,14]. Using an *in situ* current annealing technique, we find that this compound can be n-doped through Cl dissociation, drastically altering the transport behaviour from semiconducting to metallic and giving rise to superconductivity below ~9 K. This work is the first example of superconductivity in a van der Waals (vdW) superatomic crystal; more broadly, it establishes a new chemical strategy to manipulate the electronic properties of vdW materials with labile ligands.**

Superatomic crystals have generated significant interest in materials science due to their modular and hierarchical structure, which allows for the design of electronically and chemically diverse materials[3,10,14–19]. Consisting of sheets of covalently-bonded clusters, vdW superatomic crystals offer the added benefit of being exfoliatable to create two-dimensional sheets that can be further chemically modified[14]. Considering the enormous impact that atomic vdW compounds have had on fundamental science and their possible applications in nanoelectronics, energy generation and storage, catalysis and quantum computing[11,12], vdW superatomic crystals have potential for extending the scope of properties and functions available in two-dimensional materials[20].

$Re_6Se_8Cl_2$ is a vdW two-dimensional structural analogue of the Chevrel phase, itself a 3D lattice of discrete $Mo_6E_8$ (E = S, Se, Te) clusters[1,2,21]. Like the Chevrel phase, $Re_6Se_8Cl_2$ consists of an octahedral metal cluster core (Re) with each face capped by a chalcogen atom (Se). In contrast to the Chevrel phase, however, each cluster is also capped with two apical Cl atoms, which facilitate vdW stacking of the covalently bound sheets (Fig. 1a). The $Re_6Se_8$ core has been studied extensively as an isolated molecular unit[22] but until recently, little was known about the electronic properties of $Re_6Se_8Cl_2$ crystals, save for its semiconducting nature at room temperature[23] with a wide electronic band gap (1.58 eV)[13].

Chen et al. recently established that in Chevrel phase compounds, the coupling of electrons to certain intra- and inter-cluster vibrations results in superconductivity[24]. In practice, superconductivity emerges when cations are inserted into the structure to modulate the number of electrons per cluster[5–7]. Given these results, could doping also reveal superconductivity in the two-dimensional Chevrel phase analogues? A significant challenge to answering this question is the lack of a synthetic strategy to dope such materials. In this work, we show that predefined regions of thin $Re_6Se_8Cl_2$ flakes integrated into mesoscopic electrical devices can be doped with electrons by applying a novel *in situ* current annealing technique. In these regions, the carrier density is increased by 4-5 orders of magnitude and the material is transformed from a semiconductor to a metal, with superconductivity appearing below ~9 K. Outside of these regions, the material retains the characteristics of an undoped semiconductor. Structural and chemical analyses suggest that the current bias-induced doping of $Re_6Se_8Cl_2$ results from the dissociation of Cl atoms[13].

Figure 1b shows a scanning electron microscopy image of an electrical transport device. We use a pre-patterned lead geometry to electrically contact mechanically exfoliated $Re_6Se_8Cl_2$ flakes (thickness ~100 nm – 500 nm), allowing for simultaneous 4-terminal longitudinal ($R_{xx}$) and Hall resistance ($R_{xy}$) measurements (see Supplementary Sec. 1.1 for fabrication details). The $Re_6Se_8Cl_2$ flakes exhibit semiconducting behaviour with the resistance strongly increasing with decreasing temperature (Fig. 1c). Fitting to a thermally activated model, $R_{xx} = R_0 e^{\frac{(E_C-E_F)}{k_B T}}$, we determine the energy difference between the Fermi level ($E_F$) and the conduction band edge ($E_C$), $E_C-E_F$ ~ 49 meV, consistent with previous scanning tunneling spectroscopy measurements[13]. From the Hall magnetoresistance, we determine the free electronic carrier density per layer, $n_{2D}$ = $6.2 \times 10^{10}$ cm$^{-2}$ (Fig. 1d). After an *in situ* current annealing procedure, which consists of passing high bias currents through the flake (Supplementary Sec. 2.1), we observe a decrease in the sample resistance by 2-3 orders of magnitude (Fig. 1c) and an increase in the carrier density by 4-5 orders of magnitude, to $n_{2D}$ = $7.4 \times 10^{15}$ cm$^{-2}$ (Fig. 1d). The flakes subsequently exhibit metallic behaviour and a superconducting transition below ~9 K (Fig. 1c).

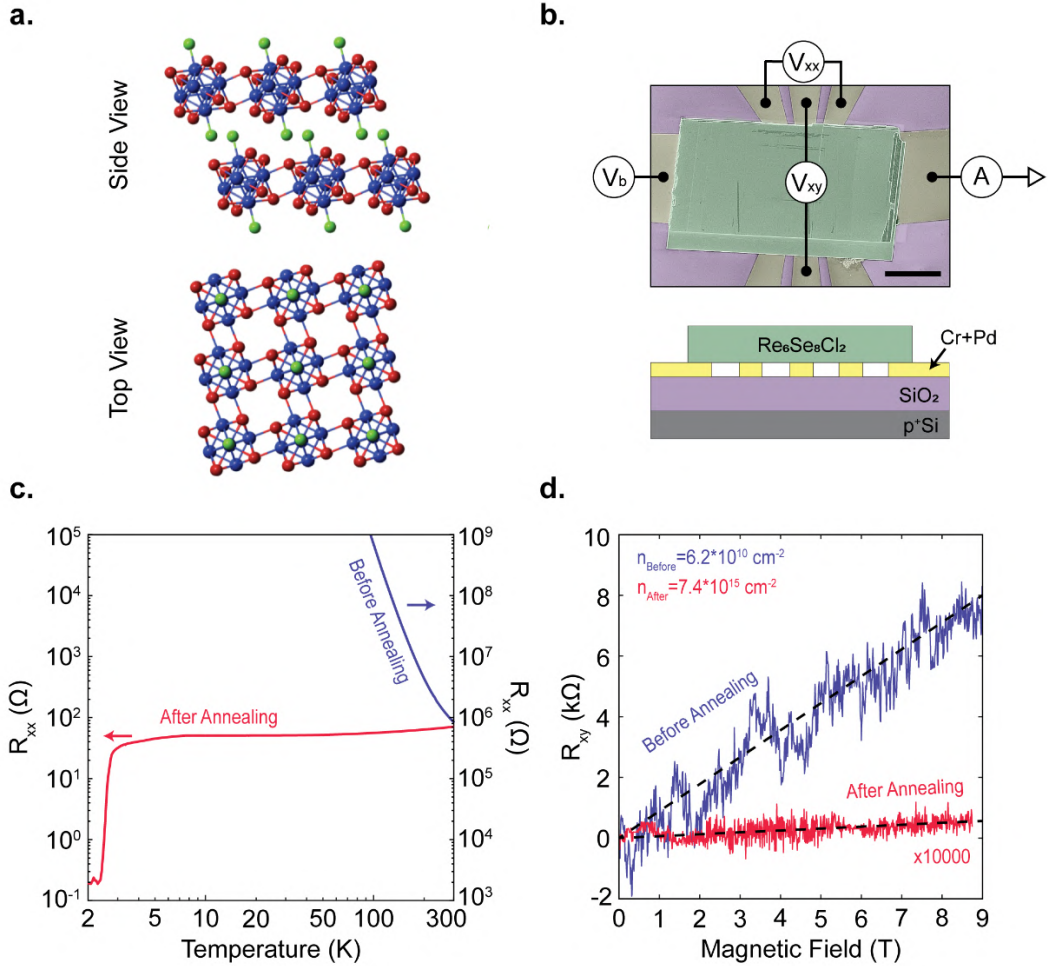

**Figure 1. Electrical transport properties of $Re_6Se_8Cl_2$. a,** Crystal structure of $Re_6Se_8Cl_2$. Colour code: Re, blue; Se, red; Cl, green. **b,** Top: SEM image of a pre-patterned transport device. The scale bar is 5 µm. The image is false coloured so the $Re_6Se_8Cl_2$ flake, pre-patterned Pd electrodes, and $SiO_2$ substrate appear green, yellow, and purple, respectively. The measurement scheme is overlaid. Bottom: Side view of the device architecture. **c,** $R_{xx}$ vs. temperature before (blue) and after (red) current annealing. **d,** Anti-symmetrized $R_{xy}$ before (blue, measured at 200 K) and after (red, measured at 10 K) current annealing. Electronic carrier density per layer is extracted using the Hall coefficient $n_{3D}*t = n_{2D} = \frac{B}{eR_{xy}} \cdot \frac{R_{xy}}{B}$ is determined by fitting the data in (**c**) to a line (dashed black lines).

Figure 2a shows a detailed measurement of the superconducting transition under varying temperature and magnetic field. At zero magnetic field, we identify three distinct regimes. At high temperature, the system appears metallic with only a weak decrease in resistance with decreasing temperature. At ~7.4 K there is an abrupt kink in the slope of the resistance, and the resistance drops more rapidly with decreasing temperature. At ~2.6 K there appears a second kink, below which the resistance drops quickly to zero, within our noise limit. We interpret the two kinks in the temperature dependence as resulting from two distinct superconducting transitions[25,26], which we label $T_C^{High}$ and $T_C^{Low}$, defined as 99% and 10% of the normal state resistance, respectively. Both $T_C^{High}$ and $T_C^{Low}$ shift to lower temperature values with increasing magnetic field (perpendicular to the cluster planes)[27]. Below $T_C^{Low}$ the resistance is thermally activated at a finite magnetic field and well fit by the Arrhenius equation $R = R_0 e^{-\frac{U(B)}{k_B T}}$, where $U(B)$ is the magnetic-field-dependent activation energy, $k_B$ is the Boltzmann constant and T is the electron temperature (see Supplementary Sec. 4.3 for details)[28]. In the intermediate regime between $T_C^{High}$ and $T_C^{Low}$, the resistance decreases approximately linearly with decreasing temperature. We have measured 8 devices fabricated in a similar way, and in all cases, we observed superconducting behaviour after current annealing (Fig. 2c). We note that while we universally observe two superconducting transitions with similar values of $T_C^{Low}$ and $T_C^{High}$ the temperature dependence in the transition region between the two superconducting transitions is sample dependent, which we interpret as a result of variability in sample inhomogeneity (Fig. 2b and Supplementary Sec. 4.1).

A full map of the field-temperature phase diagram is shown in the inset of Figure 2a. The critical magnetic field versus temperature is well fit by the relation $H_{C2} = H_{C2\text{-}0}\left(1-\left(\frac{T}{T_C}\right)^2\right)$ for both transitions, where $H_{C2}$ is determined from the $0.1 \times R_N$ and $0.99 \times R_N$ contours, $T_C$ is the corresponding zero-field critical temperature and $H_{C2\text{-}0}$ is the zero-temperature upper critical field[27]. From these fits we determine $H_{C2\text{-}0}^{Low} = 3.1$ T, and $H_{C2\text{-}0}^{High} \approx 32.3$ T. Close to $T_C$, the critical field follows a linear relation, $H_{C2} = \frac{\varphi_0}{2\pi\xi_0^2}\left(1-\frac{T}{T_C}\right)$, which we use to extract zero-temperature Ginzburg-Landau coherence lengths[27] of $\xi_0^{Low} = 10.3$ nm and $\xi_0^{High} = 3.2$ nm. Figure 2c shows an aggregate phase diagram of all measured devices confirming that the equilibrium thermodynamic properties are consistent across all samples (see also Supplementary Fig. 7).

To verify $T_C^{High}$ and $T_C^{Low}$ are consistent with superconducting transitions, we investigate the non-equilibrium properties of an annealed $Re_6Se_8Cl_2$ device (Fig. 2d-f). Figure 2d presents $R_{AC}(I_{DC}) = \frac{dV}{dI}\bigg|_{I_{DC}}$ versus DC current at 1.6 K and zero magnetic field, in which we observe two distinct peak features indicating two critical current values, $I_C^{Low} \sim$ 1.4 µA and $I_C^{High} \sim$ 4.5 µA. The magnetic field and temperature dependence of $I_C^{Low}$ and $I_C^{High}$ is shown in Figure 2e,f. The extracted thermodynamic quantities from the magnetic field and temperature dependencies[29], $T_C^{Low} = 2.9$ K, $H_{C2\text{-}0}^{Low} = 3.4$ T and $T_C^{High} = 7.0$ K, $H_{C2\text{-}0}^{High} \approx 23.5$ T (Supplementary Fig. 10) are in good agreement with the equilibrium results, confirming that $I_C^{Low}$ and $I_C^{High}$ correspond to the two transitions, $T_C^{Low}$ and $T_C^{High}$.

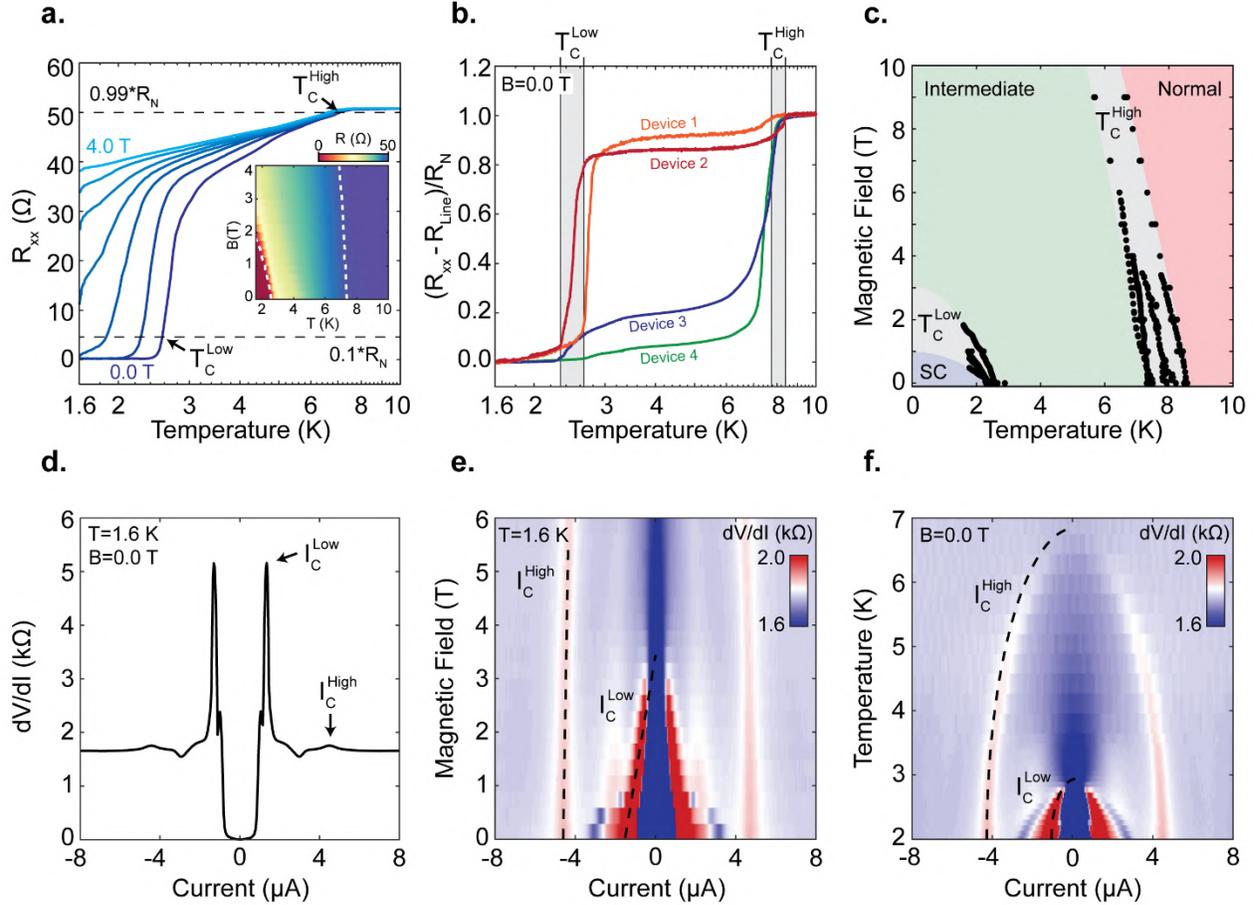

**Figure 2. Superconductivity in current annealed $Re_6Se_8Cl_2$. a,** $R_{xx}$ vs. temperature of annealed $Re_6Se_8Cl_2$ measured at various magnetic fields. Two transition temperatures ($T_C^{High} \equiv 0.99xR_N$) and ($T_C^{Low} \equiv 0.1xR_N$) are denoted. Inset: two-dimensional contour plot of $R_{xx}$ vs. temperature and magnetic field. Dashed white lines track $0.1xR_N$ and $0.99xR_N$. **b,** $R_{xx}$ vs. temperature at zero magnetic field for 4 additional devices. Each curve has a background resistance subtracted, then is normalized to the normal state resistance. The grey regions represent the ranges of $T_C^{High}$ and $T_C^{Low}$. **c,** Full magnetic field-temperature phase diagram with $T_C^{High}$ and $T_C^{Low}$ contour lines for 8 different $Re_6Se_8Cl_2$ flakes. Normal, superconducting, and intermediate regimes are denoted by red, blue, and green regions respectively. **d,** Differential resistance $R_{xx}^{AC} = \frac{dV}{dI}\big|_{I_{DC}}$ vs. DC current at 1.6 K and zero magnetic field. Two critical currents are denoted $I_C^{High}$ and $I_C^{Low}$. **e,** Two-dimensional contour plot of $R_{xx}^{AC}$ vs. DC current and magnetic field. The dashed black lines are a guide to the eye tracking $I_C^{High}$ and $I_C^{Low}$. **f,** Two-dimensional contour plot of $R_{xx}^{AC}$ vs. DC current and temperature The dashed black lines are a guide to the eye tracking $I_C^{High}$ and $I_C^{Low}$. For (**d-f**) a background resistance of 156 Ω was subtracted.

We attribute the changes in transport properties after current annealing to a loss of interplanar Cl atoms, as such vacancies have been shown to introduce mid-gap electronic states resulting in n-type doping[13,14]. The existence of two identifiable superconducting transitions in all samples suggests the existence of two distinct superconducting phases. The key result is the upper transition ($T_C^{High}$), which we associate with $Re_6Se_8$ clusters doped from Cl dissociation as the remarkably large upper critical field is comparable to other well-known superconducting 3D Chevrel phase compounds[5]. The lower superconducting transition ($T_C^{Low}$) could be a manifestation of superconducting island behavior as a result of the sample inhomogeneity[25,26] or possibly due to a formation of metallic Re[30] from cluster decomposition. While we can't discount the latter hypothesis, we note that our structural and chemical analyses did not detect the presence of metallic Re in the samples and the upper critical field for the observed transition, $H_{C2-0}^{Low}$, is ~50 times larger than metallic Re[30].

To support our proposed doping mechanism through Cl dissociation and better understand the effects of current annealing, we examine the structure and composition of current annealed flakes exhibiting superconducting behaviour. Figure 3a presents micro Raman spectra taken on the contact interface of a current annealed flake (see Supplementary Sec. 1.1 for sample preparation). We measure no deviation in the bulk Raman modes between the current annealed and non-current annealed regions[14]. We used focused ion beam (FIB) and transmission electron microscopy (TEM) to obtain a high-resolution cross-sectional image of the device structure after current annealing. We clearly see one of the pre-patterned leads and the crystal planes of the $Re_6Se_8$ cluster layers (Fig. 3b). The crystal structure is preserved up to the contact interface.

To measure the chemical state of the annealed material, we performed spatially resolved micro x-ray photoelectron spectroscopy (µ-XPS). We focus specifically on the Re 4f binding energy region to explore the possibility of forming regions of metallic Re during current annealing. As a reference, we performed XPS on $Re_6Se_8Cl_2$, both pristine (Fig. 3c) and after thermal annealing at 1000 ºC (Fig. 3d). The thermal treatment decomposes the surface of the material, and we observe the emergence of two additional peaks corresponding to metallic Re in Figure 3d (see also Supplementary Sec. 5.5.1). By comparison, the µ-XPS Re 4f spectrum of the current annealed crystal (Fig. 3e) shows no measurable signal from metallic Re. These combined structural and chemical analyses on current annealed flakes suggest the $Re_6Se_8$ clusters are preserved and any metallic Re resulting from cluster degradation is not detectable by the reported probes.

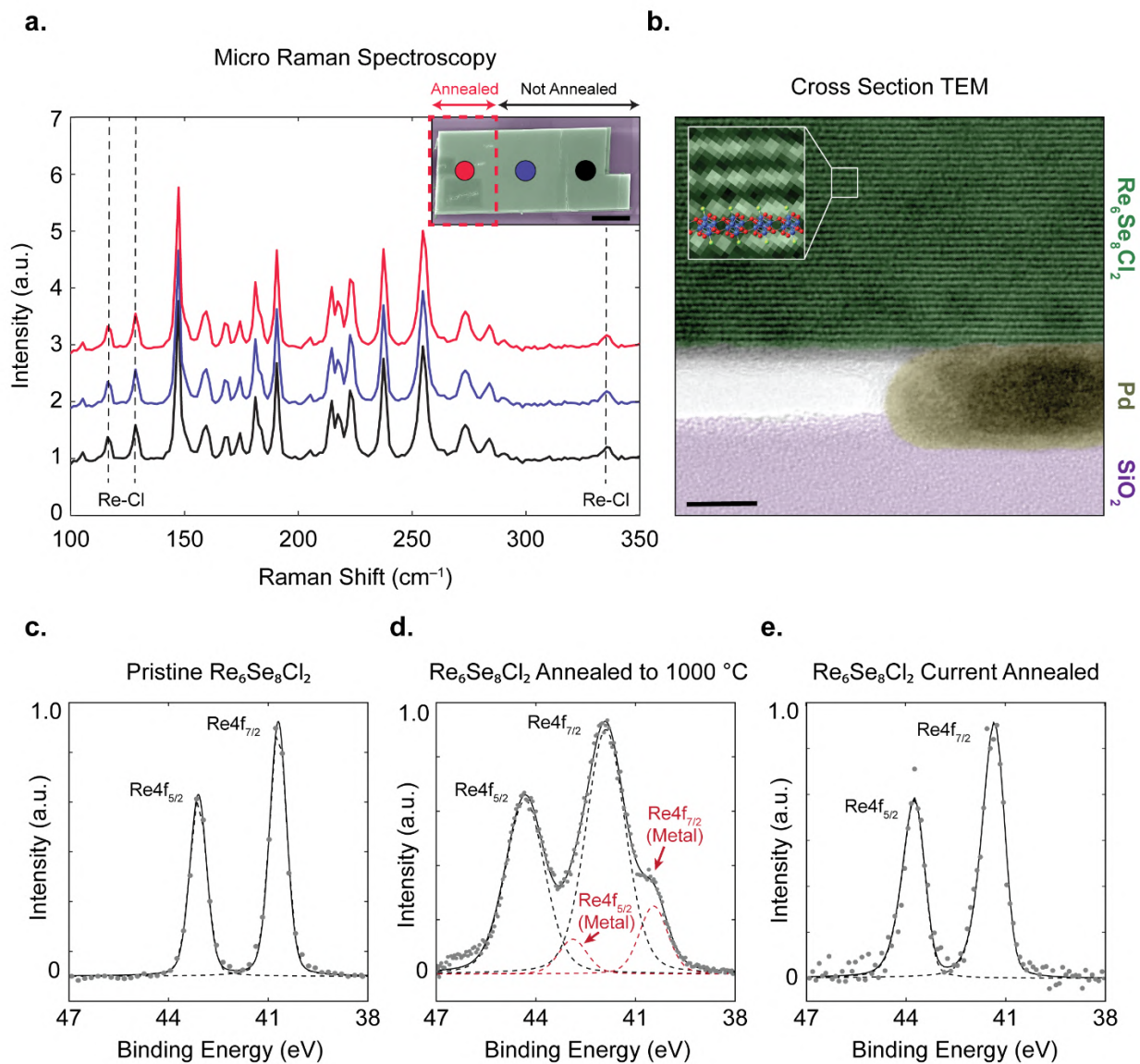

**Figure 3. Structural and chemical analyses of annealed Re₆Se₈Cl₂. a,** Microscopic Raman spectroscopy of Re₆Se₈Cl₂ after current annealing. The spectra are normalized to the intensity at 350 cm⁻¹, then offset from one another by +1. The colour of each individual spectrum corresponds to a region in the inset. Inset: SEM image of a flipped Re₆Se₈Cl₂ flake after current annealing. The scale bar is 5 μm. The image is false coloured so Re₆Se₈Cl₂ and SiO₂ appear green and purple respectively. The coloured circles mark the positions where Raman spectroscopy was acquired. The current annealed region is denoted by the red dashed box. **b,** Cross sectional TEM image of a current annealed flake. The scale bar is 10 nm. The image is false coloured so Re₆Se₈Cl₂, SiO₂, and Pd are green, purple, and yellow respectively. Inset: enlarged section of the TEM image identifying the Re₆Se₈ cluster planes. **c,** μ-XPS Re 4f spectra of pristine micro crystalline Re₆Se₈Cl₂. **d,** XPS Re 4f spectra of micro crystalline Re₆Se₈Cl₂ thermally annealed at 1000 °C. **e,** Re 4f μ-XPS spectrum on the current annealed Re₆Se₈Cl₂ flake shown in (**a**). For (**c**), (**d**), and (**e**) the dashed black and red lines represent single Gaussian fits to the data. Solid black lines are the total fit.

The exact mechanism responsible for doping of the $Re_6Se_8Cl_2$ crystals remains an open question. Our hypothesis is that localized heating during the current annealing process leads to the dissociation of Cl from the sheets. To support this, we performed three-dimensional steady state simulations of the heat transfer process using ANSYS Workbench (see Supplementary Sec. 3 for details). Our simulations suggest the temperature can exceed the Cl dissociation temperature (Fig. 4), with a strongly localized spatial dependence.

If Cl dissociation does indeed occur due to localized heating, we expect the behaviour to be reproducible through a purely thermal process. To understand the relationship between annealing temperature, elemental composition and electrical properties, we subjected $Re_6Se_8Cl_2$ to a series of heat treatments and characterized the resulting samples. For these experiments, the material is sealed in fused silica tubes under vacuum and annealed at set temperatures up to 1000 ºC. We use XPS and energy dispersive x-ray (EDX) spectroscopy to determine the surface and bulk elemental compositions, respectively. When the material is heated to 400 ºC, XPS reveals a decrease in the surface Cl concentration when compared to the pristine material (Fig. 4c), while EDX only shows a slight decrease in the bulk Cl signal. Between 400 and 600 ºC, the bulk Cl concentration shows a dramatic decrease (Fig. 4b), which coincides with a decrease in the resistance of the material by 1-2 orders of magnitude (Fig. 4a). When $Re_6Se_8Cl_2$ is annealed to 1000 ºC, we observe a further decrease of the bulk Cl elemental composition by EDX, as well as a loss of Se, indicating an onset of cluster decomposition. We note that single crystal x-ray diffraction indicates that the bulk crystal structure is unchanged across the whole temperature range (Supplementary Sec. 5.4).

Our observations of Cl loss are corroborated by performing real time thermal annealing - mass spectrometry (TA-MS) analysis (Supplementary Sec. 5.7). Figure 4d shows the combined intensity responses for the $Cl^-$ and $Cl_2^-$ ions when $Re_6Se_8Cl_2$ (green) and a control empty crucible (black) are gradually heated from 25 to 800 °C, while the red trace measures the combined $Se^-$, $Se_2^-$ and $Se_3^-$ signals. The Cl signal shows two peaks at ~385 and ~763 °C, suggesting the onset of two different Cl dissociation processes. The higher temperature peak agrees with the bulk Cl loss temperature while the lower temperature peak may be attributed to the dissociation of surface level Cl atoms. We also note that no Se signal is observed up to 800 °C. Taken together, these results demonstrate that the Cl dissociation and doping process can be thermally driven and occur without a loss of the intrinsic $Re_6Se_8$ cluster structure.

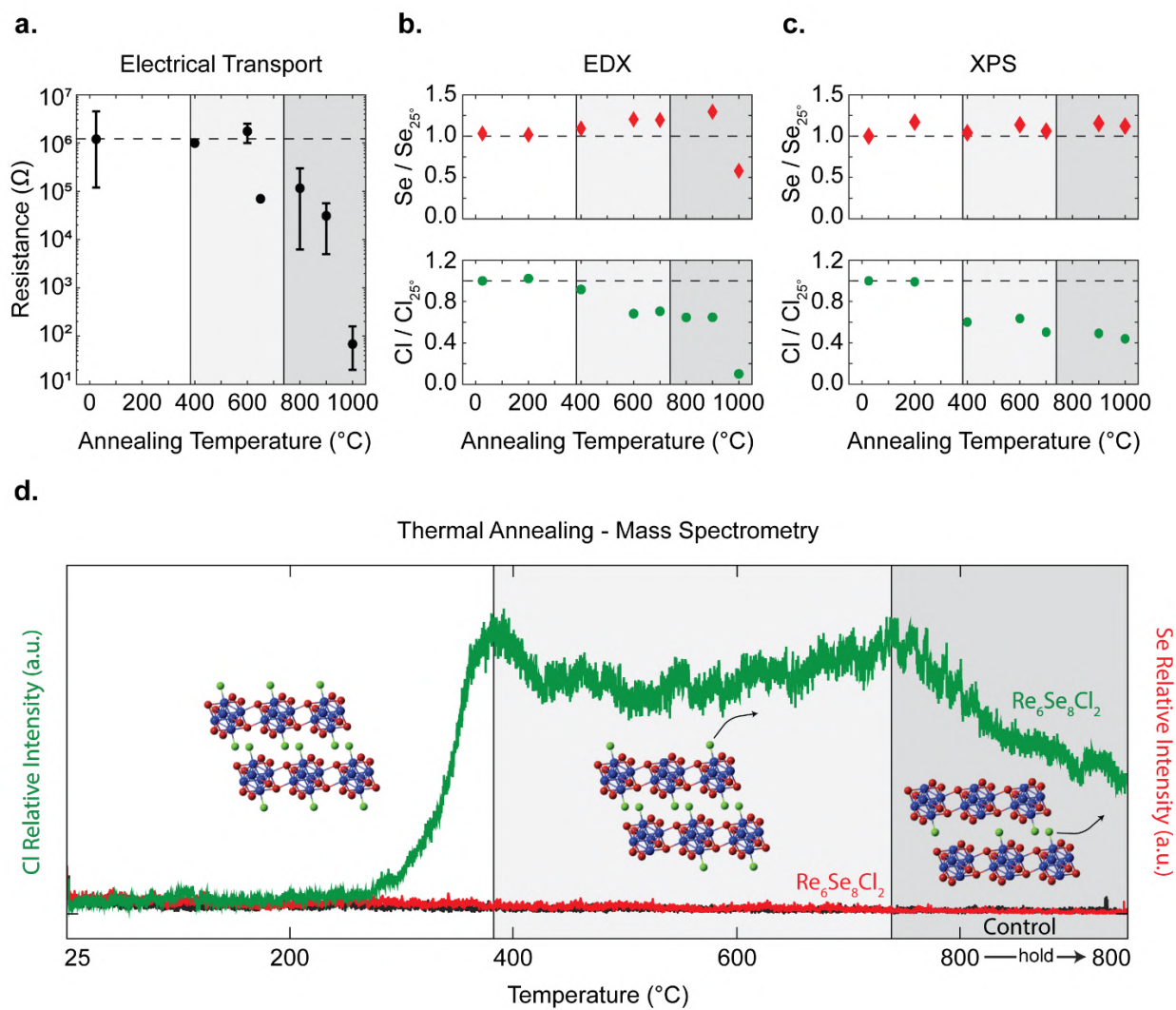

**Figure 4. Chemical analysis of thermally annealed Re$_6$Se$_8$Cl$_2$. a,** Two-terminal resistance of bulk single crystal Re$_6$Se$_8$Cl$_2$ vs. annealing temperature. Resistance was measured in the linear IV regime. **b,** Relative Se and Cl loss measured by EDX versus annealing temperature. Values for each element were calculated by dividing the measured atomic percentage at each temperature by the atomic percentage of the non-annealed sample (25 °C). **c,** Relative Se and Cl loss measured by XPS vs. annealing temperature. Values were calculated by dividing each element's integrated spectrum area by the Re 4f spectrum area for each temperature, then dividing each ratio by the non-annealed ratio (25 °C). **d,** Cl$_2^-$ and Cl$^-$ mass counts vs. time as Re$_6$Se$_8$Cl$_2$ powder (green) and an empty crucible (black) are heated from 25 °C to 800 °C. The red line is a measurement of Se$^-$, Se$_2^-$, and Se$_3^-$ counts for the Re$_6$Se$_8$Cl$_2$ powder. In (**a**) through (**d**), the white, light grey, and dark grey regions correspond to no loss of Cl, a loss of surface Cl, and a loss of bulk Cl, respectively. A schematic of the proposed dissociation process is shown for each region in (**d**).

The bulk thermal annealing results, in combination with the thermal modelling and the chemical and structural analysis on current annealed flakes, suggest that the current annealing procedure results in electron doping through Cl dissociation as a result of sample heating. While thermal annealing has not reproduced the same signatures of superconductivity, future work will focus on exploring the vast thermal annealing phase space to produce bulk $Re_6Se_8$ cluster superconductivity.

We have reported the first observation of superconductivity in a vdW superatomic crystal. Through both current and thermal annealing processes, we demonstrate that $Re_6Se_8Cl_2$ can lose interplanar Cl atoms, while retaining the integrity of the $Re_6Se_8$ clusters. This leads to a dramatic increase in carrier concentration and gives rise to clear signatures of robust $Re_6Se_8$ cluster superconductivity. These results offer the possibility of generalizing this approach to electronically dope other vdW materials with labile halide ligands. We envision using the annealing process to write metallic or superconducting patterns on the nanoscale, using techniques such as atomic force microscopy and scanning tunneling spectroscopy to controllably dope well-defined regions.

**Acknowledgements**

We thank Abhay Pasupathy, James Hone, and Avishai Benyamini for fruitful discussions and input. We also thank Dan Paley for assistance with the SCXRD and XPS measurements, and Tai-De Li for assistance with the micro XPS. Sample preparation and transport measurements were primarily supported by the NSF MRSEC program through the Center for Precision Assembly of Superstratic and Superatomic Solids (DMR-1420634). Chemical and structural analyses (XPS, TA-MS, SCXRD, TEM) were supported by the NSF CAREER award (DMR-1751949). Raman spectroscopy was supported by the US Air Force Office of Scientific Research (AFOSR) grant FA9550-18-1-0020. Thermal transport modeling was supported by the Army Research Office grant ARO#71641-MS. J.C.R. is supported by the U.S. Department of Defense through the National Defense Science & Engineering Graduate Fellowship (NDSEG) Program.

**Author Contributions**

EJT, JR, XR, and CRD designed the experiments. EJT, JR, XR, and CRD analyzed the data. EJT and JRS fabricated pre-patterned transport devices. JR fabricated bulk transport devices. EJT and JRS performed the transport measurements and current annealing procedures. JR synthesized $Re_6Se_8Cl_2$ crystals. JR performed the bulk crystal thermal annealing and corresponding XPS, SCXRD, and EDX measurements. JR and BF performed the TA-MS experiment. EJT and KL performed Raman spectroscopy measurements. AZ performed the FIB slice and TEM measurements. XW and JAM performed bulk thermal conductivity measurements and simulated the temperature rise for the current annealing procedure. All authors discussed the data and contributed to writing the manuscript.

**Competing Interests**

The authors declare no competing interests.

**Corresponding Authors**

Xavier Roy – xr2114@columbia.edu

Cory R. Dean – cd2478@columbia.edu

**References:**

1.  Burdett, J. K. & Lin, J. H. The structures of Chevrel phases. *Inorg. Chem.* **21**, 5–10 (1982).

2.  Peña, O. Chevrel phases: Past, present and future. *Phys. C Supercond. its Appl.* **514**, 95–112 (2015).

3.  Pinkard, A., Champsaur, A. M. & Roy, X. Molecular Clusters: Nanoscale Building Blocks for Solid-State Materials. *Acc. Chem. Res.* **51**, 919–929 (2018).

4.  Reber, A. C. & Khanna, S. N. Superatoms: Electronic and Geometric Effects on Reactivity. *Acc. Chem. Res.* **50**, 255–263 (2017).

5.  Fischer, Ø. Chevrel phases: Superconducting and normal state properties. *Appl. Phys.* **16**, 1–28 (1978).

6.  Xie, W. *et al.* Endohedral gallide cluster superconductors and superconductivity in ReGa5. *Proc. Natl. Acad. Sci. U. S. A.* **112**, E7048-54 (2015).

7.  Ganin, A. Y. *et al.* Polymorphism control of superconductivity and magnetism in Cs3C60 close to the Mott transition. *Nature* **466**, 221–225 (2010).

8.  Lee, C.-H. *et al.* Ferromagnetic Ordering in Superatomic Solids. *J. Am. Chem. Soc.* **136**, 16926–16931 (2014).

9.  Allemand, P. M. *et al.* Organic molecular soft ferromagnetism in a fullerene c60. *Science* **253**, 301–2 (1991).

10. Roy, X. *et al.* Nanoscale Atoms in Solid-State Chemistry. *Science (80-. ).* **341**, 157–160 (2013).

11. Geim, A. K. & Grigorieva, I. V. Van der Waals heterostructures. *Nature* **499**, 419–425 (2013).

12. Novoselov, K. S., Mishchenko, A., Carvalho, A. & Neto, A. H. C. 2D materials and van der Waals heterostructures. *Science (80-. ).* **353**, aac9439 (2016).

13. Zhong, X. *et al.* Superatomic Two-Dimensional Semiconductor. *Nano Lett.* **18**, 1483–1488 (2018).

14. Choi, B. *et al.* Two-Dimensional Hierarchical Semiconductor with Addressable Surfaces. *J. Am. Chem. Soc* **140**, 9369–9373 (2018).

15. Hernández Sánchez, R. *et al.* Electron Cartography in Clusters. *Angew. Chemie Int. Ed.* **57**, 13815–13820 (2018).

16. Champsaur, A. M. *et al.* Building Diatomic and Triatomic Superatom Molecules. *Nano Lett.* **16**, 5273–5277 (2016).

17. O'Brien, E. S. *et al.* Single-crystal-to-single-crystal intercalation of a low-bandgap superatomic crystal. *Nat. Chem.* **9**, 1170–1174 (2017).

18. Ong, W.-L. *et al.* Orientational order controls crystalline and amorphous thermal transport in superatomic crystals. *Nat. Mater.* **16**, 83–88 (2017).

19. Lee, K. *et al.* Two-Dimensional Fullerene Assembly from an Exfoliated van der Waals Template. *Angew. Chemie Int. Ed.* **57**, 6125–6129 (2018).

20. Choi, B. *et al.* Van der Waals Solids from Self-Assembled Nanoscale Building Blocks. *Nano Lett.* **16**, 1445–1449 (2016).

21. Leduc, L., Perrin, A. & Sergent, M. Structure du Dichlorure et Oetaseleniure

d'Hexarhenium, Re6Se8Cl2: Compose Bidimensionnel a Clusters Octaedriques Re6. *Acta Crystallogr.* **C39**, 1503–1506 (1983).

22. Gabriel, J.-C. P., Boubekeur, K., Uriel, S. & Batail, P. Chemistry of Hexanuclear Rhenium Chalcohalide Clusters. *Chem. Rev.* **101**, 2037–2066 (2001).

23. Leduc, L., Padiou, J., Perrin, A. & Sergent, M. Synthese et caracterisation d'un nouveau chalcohalogenure a clusters octahedriques de rhenium a caractere bidimensionnel: Re6Se8Cl2. *J. Less-Common Met.* **95**, 73–80 (1983).

24. Chen, J., Millis, A. J. & Reichman, D. R. Intermolecular coupling and superconductivity in PbMo6S8 and other Chevrel phase compounds. *Phys. Rev. Mater.* **2**, 114801 (2018).

25. Eley, S., Gopalakrishnan, S., Goldbart, P. M. & Mason, N. Approaching zero-temperature metallic states in mesoscopic superconductor–normal–superconductor arrays. *Nat. Phys.* **8**, 59–62 (2012).

26. Sun, Y. *et al.* Double quantum criticality in superconducting tin arrays-graphene hybrid. *Nat. Commun.* **9**, 2159 (2018).

27. Tinkham, M. *Introduction to superconductivity. Dover books on physics* (Dover Publ, 2004).

28. Feigel'man, M. V., Geshkenbein, V. B. & Larkin, A. I. Pinning and creep in layered superconductors. *Phys. C Supercond.* **167**, 177–187 (1990).

29. Dew-Hughes, D. The critical current of superconductors: an historical review. *Low Temp. Phys.* **27**, 713–722 (2001).

30. Daunt, J. G. & Smith, T. S. Superconductivity of Rhenium. *Phys. Rev.* **88**, 309–311 (1952).

**Methods**

**Single crystal Re$_6$Se$_8$Cl$_2$ synthesis**

Re (330 mg, 1.80 mmol), Se (190 mg, 2.41 mmol), and ReCl$_5$ (200 mg, 0.55 mmol) are ground with a mortar and pestle and pressed into a pellet under N$_2$ atmosphere. The pellet is sealed in a quartz tube of approximately 30 cm in length under vacuum. Large single crystals are grown by chemical vapor transport method, with ReCl$_3$ acting as the transport agent. The tube is heated in a tube furnace to 1100 °C at 1 °C/min and held at 1100 °C for 3 days. With the pellet in the hot end, the tube is cooled to a 970 - 925 °C gradient over 7 hours and held for 200 hours. The tube is further cooled to a 340 - 295 °C gradient over 100 hours, and the furnace is shut off. Large 0.5 mm single crystals are recovered from the middle zone of the tube. Excess ReCl$_3$ is removed from the single crystals by using a gradient of 300 °C - RT to condense the liquid ReCl$_3$ to the cooler end.

**Transport measurements**

Before current annealing, transport measurements are performed in a voltage bias scheme, where a lock-in amplifier sources an AC voltage (at a frequency of 17.777 Hz) and auxiliary lock-in amplifiers measure the AC current, 4-probe longitudinal voltage, and Hall voltage. After current annealing, measurements are performed in a current bias scheme where a large resistor (1-10 MΩ) is placed in series before the sample to limit the AC current to ~100 nA. In both schemes, the longitudinal resistance $R_{xx}$ is defined as $\frac{V_{xx-AC}}{I_{AC}}$ and the hall resistance $R_{xy}$ is defined as $\frac{V_{xy-AC}}{I_{AC}}$. For all equilibrium transport measurements, the DC voltage and DC current are zero. For non-equilibrium measurements, a DC+AC current bias scheme is used where a lock-in amplifier is connected in parallel with a DC voltage source. A 10 MΩ resistor is placed after the lock-in output and a 100 kΩ resistor is placed after the DC voltage source output. The AC current is measured with the source lock-in and the DC current is measured with the DC voltage source. The resistance is defined as $R(I_{DC}) = \frac{dV}{dI}\big|_{I_{DC}} = \frac{V_{xx-AC}}{I_{AC}}$. An AC current of ~100 nA is used.

Bulk transport measurements are performed in a two-terminal configuration where a lock-in amplifier sources AC voltage and measures AC current (at a frequency of 17.777 Hz). The bulk resistance is defined as $R = \frac{V_{source-AC}}{I_{AC}}$, which includes the line and contact resistance (which is small compared to the sample resistance).

**Data Availability:**

The data that support the findings of this study are available from the corresponding author upon reasonable request.

**References:**

31. Novoselov, K. S. *et al.* Electric field effect in atomically thin carbon films. *Science* **306**, 666–9 (2004).
32. Wang, L. *et al.* One-Dimensional Electrical Contact to a Two-Dimensional Material. *Science* **342**, 614–617 (2013).

**Supplementary Information**

**Doping-induced superconductivity in the van der Waals superatomic crystal Re$_6$Se$_8$Cl$_2$**

## 1. Transport device fabrication

### 1.1 Re$_6$Se$_8$Cl$_2$ flake devices

Pre-patterned metal electrodes are fabricated on a p$^+$Si/SiO$_2$ (285 nm) substrate using electron beam lithography and deposition to define the device geometry. The electrodes are fabricated in two steps. The first defines the Re$_6$Se$_8$Cl$_2$ flake contact area, consisting of Cr + Pd with a thickness of 2 nm + 10 nm respectively. The second defines the wire bonding pads, which consist of Cr + Pd + Au with a thickness of 2 nm + 40 nm + 50 nm respectively. Electrodes are cleaned before use with a low power O$_2$ plasma etch for 5 minutes and screened for cleanliness and homogeneity with an atomic force microscope. Thin Re$_6$Se$_8$Cl$_2$ flakes (100 nm – 500 nm) are prepared by mechanical exfoliation with Scotch tape[1], then transferred onto a polydimethylsiloxane (PDMS) substrate. Once on the PDMS, flakes are optically identified and deterministically transferred onto the pre-patterned electrodes using the dry polymer transfer technique[2] (Supplementary Fig. 1). The substrate is then glued to a DIP socket with conducting epoxy (SPI silver paste plus) and wire bonded for transport measurements.

To perform chemical and structural analyses on the current annealed flakes, we first expose the Re$_6$Se$_8$Cl$_2$/electrode interface using the flip-chip transfer technique. Distinct metallic features are then deposited in the vicinity of the flake using electron beam lithography and deposition to help identify the sample region for micro x-ray photoemission spectroscopy (µ-XPS), energy dispersive x-ray spectroscopy (EDX) and micro Raman spectroscopy (Supplementary Fig. 1).

**1.2 Bulk single crystal Re$_6$Se$_8$Cl$_2$ devices**

Bulk synthesized Re$_6$Se$_8$Cl$_2$ crystals are bonded to a DIP socket using a low temperature non-conducting epoxy (Loctite EA 1C). Connections to the sample are made directly with silver paint (Dupont 4929N).

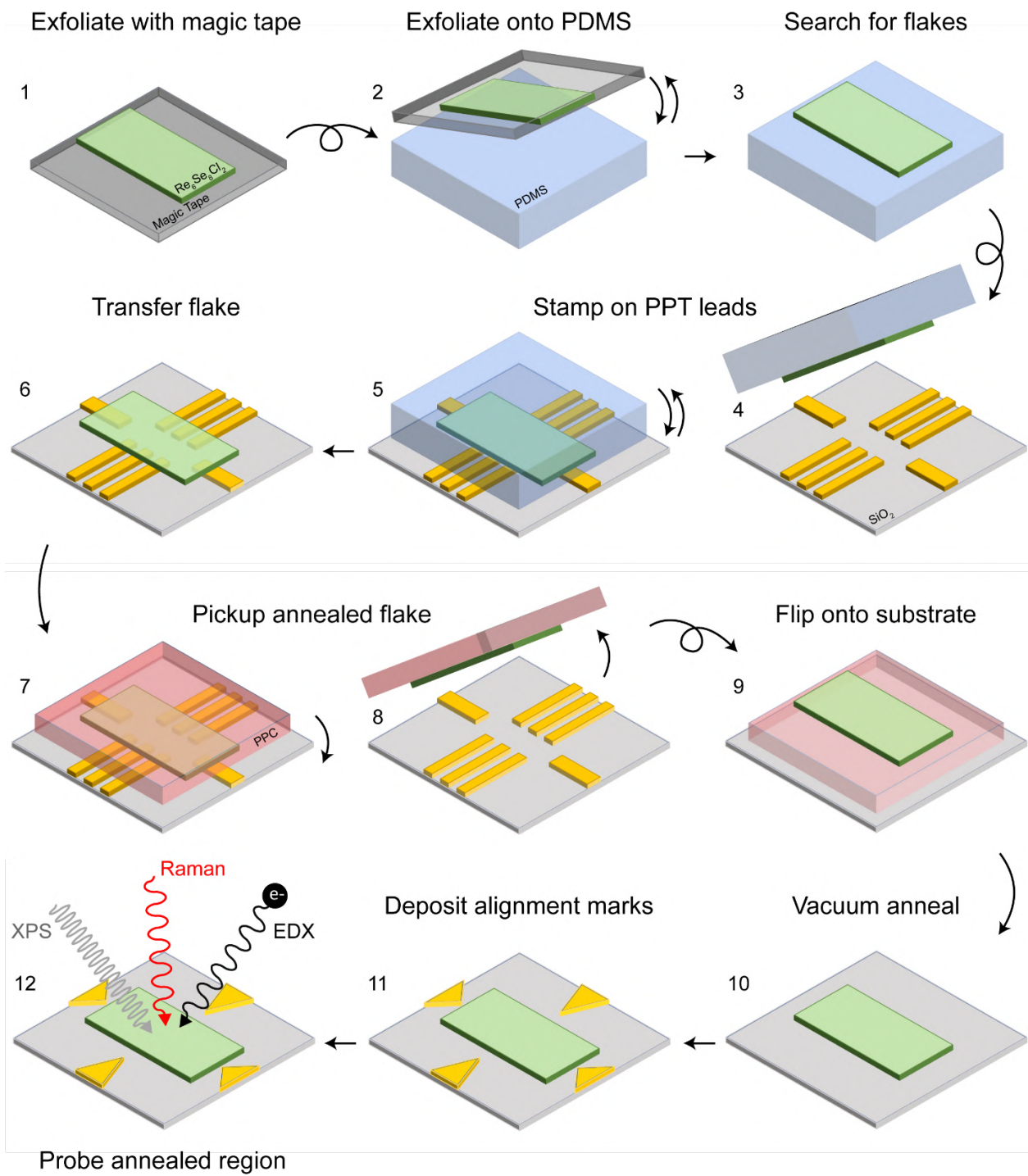

**Supplementary Figure 1. $Re_6Se_8Cl_2$ device fabrication and analysis.** Process flow for fabrication of pre-patterned $Re_6Se_8Cl_2$ transport devices and analysis of the current annealed region.

## 2. Current annealing procedure

### 2.1 Pre-patterned devices

We source DC voltage and measure DC current in a 2-terminal configuration. To protect the samples from current spikes, we set a current compliance to limit the maximum current through the sample and add a 100 kΩ resistor in series. We then ramp the source voltage incrementally until the compliance current is met. The voltage is then ramped down to zero after which we measure the sample resistance with low biases. This process is repeated, incrementally increasing the compliance current (thus the maximum power passed through the sample) until the sample resistance decreases significantly (<10 kΩ) and saturates (Supplementary Fig. 2). We anneal the shortest sample channels sequentially to produce a longer sample path that exhibits metallic behavior. For example, in Supplementary Figure 2b, we would anneal channels 1-4 in order, which produces a longer metallic sample that can be measured in a 4-terminal voltage configuration. Supplementary Figure 2b presents the measured sample resistance vs. annealing current for a $Re_6Se_8Cl_2$ flake. Empirically, we take ~$10^{-4}$ A as an upper limit. This corresponds to an input electrical power of ~2-30 mW (range is for all samples). The entire procedure is done at 1.6 K.

### 2.2 Bulk single crystal samples

The current annealing procedure is identical for bulk single crystal $Re_6Se_8Cl_2$ samples except the annealing was done at 200 K and the upper limit on current is taken as ~$10^{-1}$ A. This corresponds an input electrical power of ~2-3 W. Supplementary Figure 2c presents the measured sample resistance vs. annealing current for a few single crystal $Re_6Se_8Cl_2$ samples.

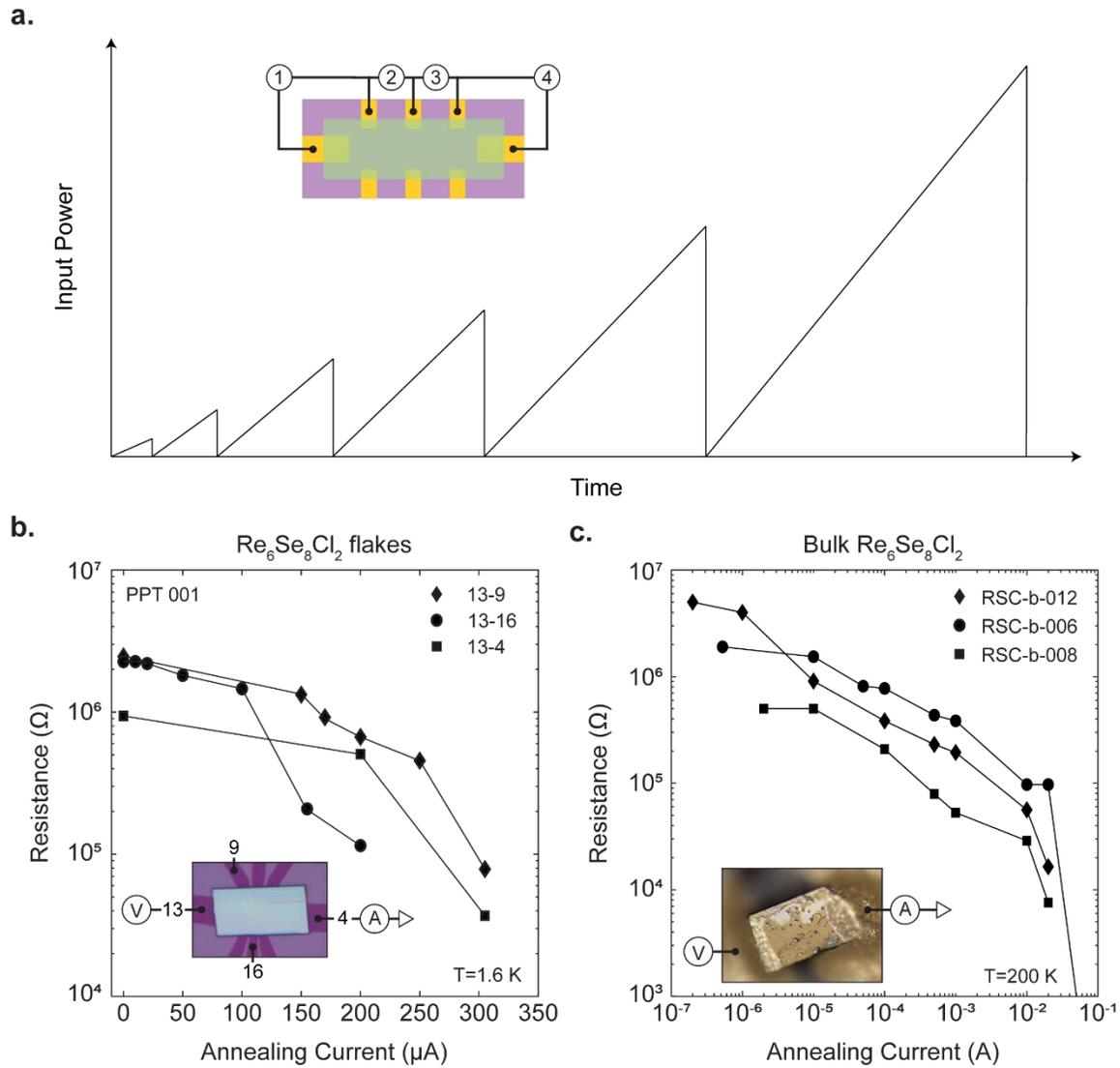

**Supplementary Figure 2. Re$_6$Se$_8$Cl$_2$ current annealing procedure. a,** Schematic of the current annealing procedure. Inset presents the current annealing geometries. Each number refers to a 2-terminal voltage bias configuration. **b,** Sample resistance vs. annealing current for a pre-patterned Re$_6$Se$_8$Cl$_2$ device. Inset presents the annealing geometries. **c,** Sample resistance vs. annealing current for three bulk single crystal Re$_6$Se$_8$Cl$_2$ devices. Inset shows the annealing geometry. For (**a**) and (**b**), resistances were measured in the low bias regime.

## 3. Modelling temperature rise during current annealing

### 3.1 Simulation details

To estimate the temperature rise due to current annealing, a steady-state thermal simulation was performed in ANSYS Workbench 18.1. The three-dimensional model with lateral dimensions based on the device in Figure 1b was built using SpaceClaim. The cross-plane thicknesses were 300 nm of $Re_6Se_8Cl_2$, 12 nm of Pd, 285 nm of $SiO_2$, and 1 µm of Si. Temperature dependent thermal conductivities of Si, $SiO_2$ and Pd were used in the simulation[3–7]. The thermal conductivity of $Re_6Se_8Cl_2$ was independently measured using Frequency Domain Thermoreflectance[8]. An in-plane thermal conductivity of 7.5 ± 1.2 W/m-K and a cross-plane thermal conductivity of 1.2 ± 0.2 W/m-K were obtained at 300 K and assumed to be independent of temperature in the simulation. Predictions suggest that current annealing brings the $Re_6Se_8Cl_2$ temperature above 300 K where similar superatomic crystals have exhibited temperature independent thermal conductivity[9]. Interface thermal conductance values were set to be 50 MW/m$^2$-K.

The spatial extent of heat deposition by joule heating is unknown and hence two heated spot sizes, shown in Supplementary Figure 3, were simulated. Supplementary Figure 3a considers a larger lateral region between the biased electrodes, while Supplementary Figure 3b considers the smallest possible lateral region. Heat was deposited on the bottom surface of the $Re_6Se_8Cl_2$ where current densities are expected to be highest. The simulations were run at steady state because the timescale for transient heat transfer is far shorter than the annealing process: less that 1 s due to the small size of the heat source.

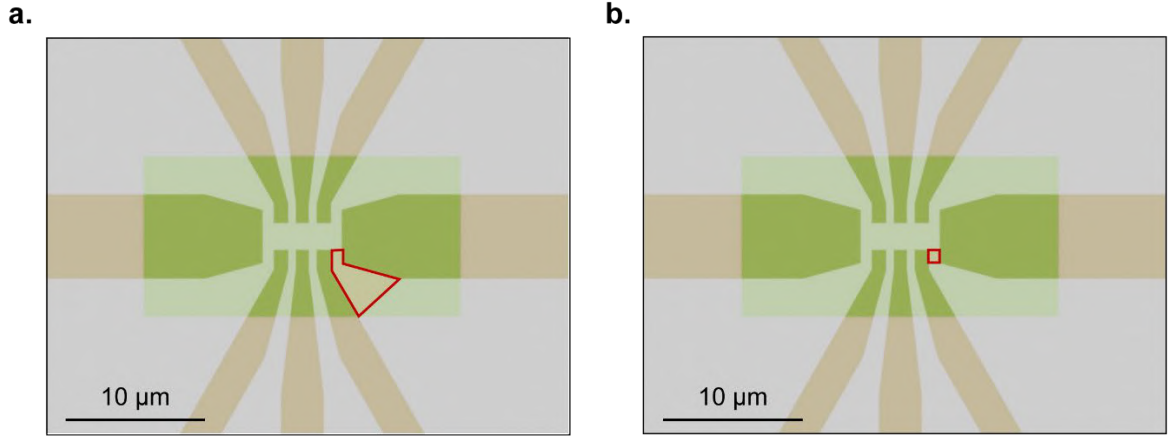

**Supplementary Figure 3. Top down schematic of the simulation domain.** Pd electrodes are beige and $Re_6Se_8Cl_2$ is transparent green to clarify the electrode positions. The two sizes of heat spot are outlined in red. **a,** Setup A, representing the upper limit of heat spot size in the annealing process. **b,** Setup B, representing the lower limit of heat spot size in the current annealing process.

### 3.2 Boundary conditions

1. Our truncated domain meets the macroscopic Si substrate at the Si bottom and side boundaries. Here, the conduction shape factor for a disk (representing our simulation domain) in a semi-infinite medium (representing the macroscopic substrate) was used to establish the convection style heat transfer boundary condition (i.e. q" = $Sk(T-T_\infty)/A$ where S is the shape factor, k is the thermal conductivity of the Si, A is the area of the boundary, and $T-T_\infty$ is the temperature difference between the boundary nodes and the cryostat base temperature)[10].

2. In the region without Pd electrodes underneath, a 12 nm gap exists between the bottom of $Re_6Se_8Cl_2$ and the top of the $SiO_2$. Heat transport across this gap is based on kinetic theory of He, the monatomic gas. Thermal conductivity k = $\frac{1}{3}c_v v \Lambda$, with a mean free path $\Lambda$ = 12 nm (because intrinsic mean free paths of gaseous He exceed the gap size), volumetric heat capacity $c_v = \frac{1}{3}k_B \eta$ where the number density η is calculated from the ideal gas law, and average speed $v = \sqrt{3k_B T/m}$ from the Maxwell Speed Distribution for molecules of mass m.

3. All other exposed surfaces of $SiO_2$, Pd and $Re_6Se_8Cl_2$ were set to have a convection heat transfer with flowing gaseous He. The convection heat transfer coefficient was calculated using the average Nusselt number $\overline{Nu_L} = 0.664\overline{Re_L}^{1/2}Pr^{1/3}$ for laminar flow over a flat plate. A flow velocity of 6.8 mm/s (based on He flow measurements) and length of 100 μm were used for $\overline{Re_L}$. Viscosity, thermal conductivity, and Prandtl number were set based on the film temperature using values from reference[10].

**3.3 Results**

The temperature map with heat source A is given in Supplementary Figure 4, and that with heat source B is given in Supplementary Figure 5. The results of the maximum temperature rise are summarized in Supplementary Table 1. The temperature maps indicate that the maximum temperature occurs at the bottom surface of the $Re_6Se_8Cl_2$ coincident with the heat source. The 12 nm gaseous He gap is a large thermal resistance that prohibits heat transfer directly into the underlying substrate, and leads to a large temperature jump shown in Supplementary Figures 4 and 5. Instead, the heat spreads laterally within the $Re_6Se_8Cl_2$ and then transfers into the substrate across the Pd contacts. The temperature rise is not linear with the magnitude of the heat input due to temperature dependent material properties. Reducing the heat source size has a significant influence on the temperature rise, and the maximum possible temperature rise exceeds 4500 K assuming the minimum possible heat source size and the maximum heat input.

| Max Temperature Rise (K) | Heat Source A | Heat Source B |
|---|---|---|
| 2 mW Heat Input | 116.9 | 362.1 |
| 30 mW Heat Input | 1089.3 | 4579.4 |

**Supplementary Table 1. Maximum temperature rise due to current annealing**

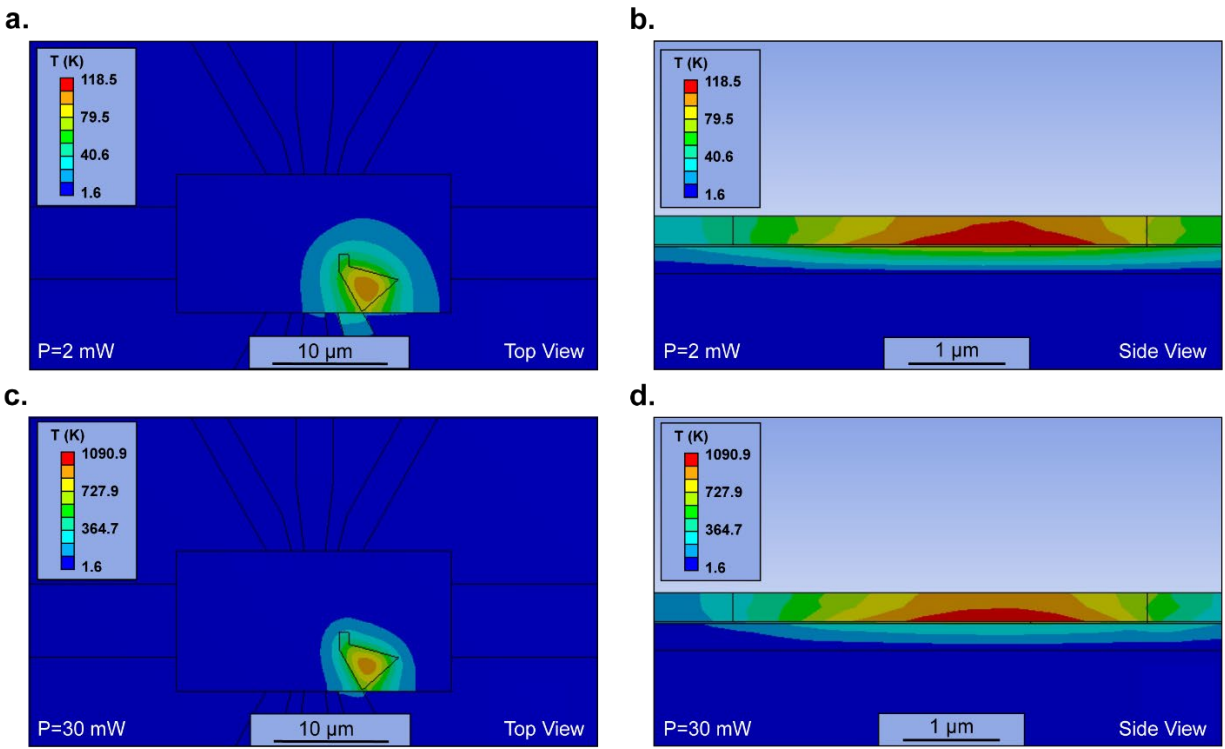

**Supplementary Figure 4. Temperature map of heat spot with setup A.** 2 mW heat is input in (**a**) and (**b**), and 30 mW heat is input for (**c**) and (**d**). (**a**) and (**c**) are the top view for the temperature map where the heat source is outlined in thin black. (**b**) and (**d**) show the cross-sectional view of the temperature distribution at the heat source.

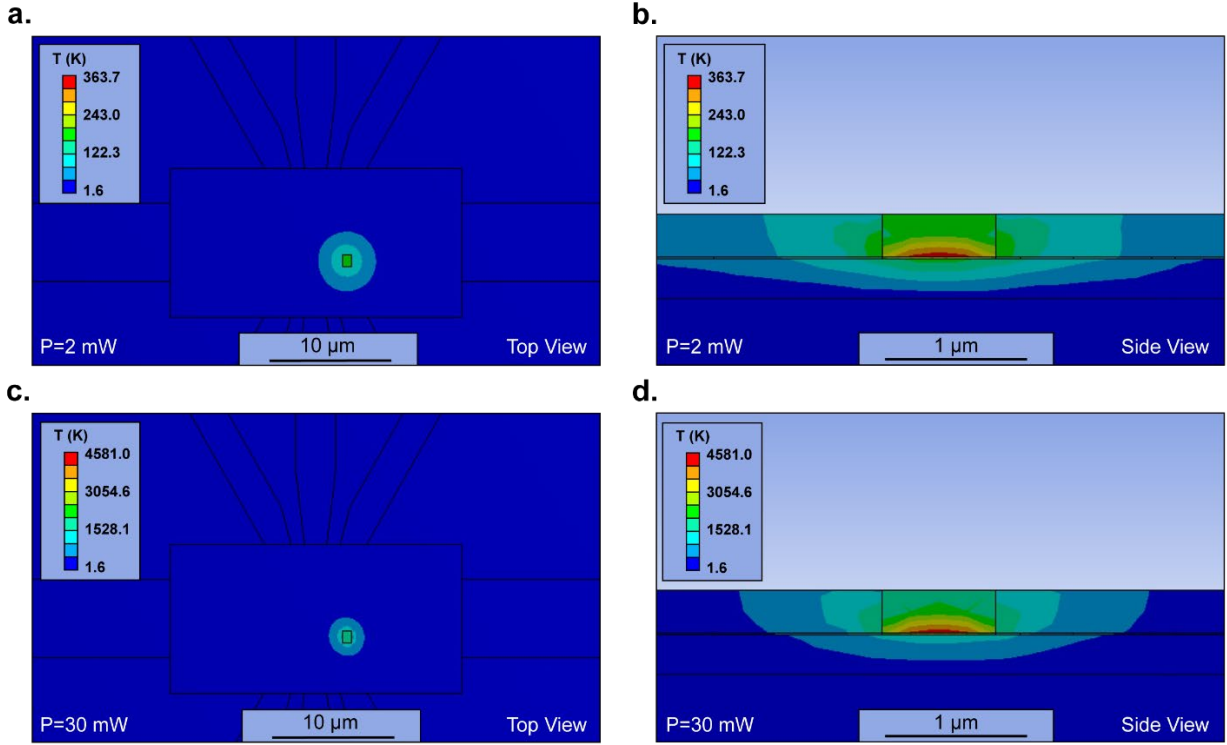

**Supplementary Figure 5. Temperature map of heat spot with setup B.** 2 mW heat is input in (**a**) and (**b**), and 30 mW heat is input for (**c**) and (**d**). (**a**) and (**c**) are the top view for the temperature map where the heat source is outlined in thin black. (**b**) and (**d**) show the cross-sectional view of the temperature distribution at the heat source.

## 4. Details of annealed $Re_6Se_8Cl_2$ superconducting properties

### 4.1 Variability in superconducting transitions

In Supplementary Figure 6, we present resistance measurements vs. temperature and magnetic field for 3 devices, whose properties are representative of the entire sample set. Each device shows two clear kinks in the resistance corresponding to $T_C^{High}$ and $T_C^{Low}$ (Supplementary Fig. 6a,b,c) and corresponding peaks in non-equilibrium measurements, $I_C^{High}$ and $I_C^{Low}$ (Supplementary Fig. 6d,e,f). Most devices exhibit multiple peaks for both $I_C^{High}$ and $I_C^{Low}$, which we attribute to sample inhomogeneity. The multiple $I_C^{High}$ and $I_C^{Low}$ peaks were assigned based on their corresponding temperature dependence.

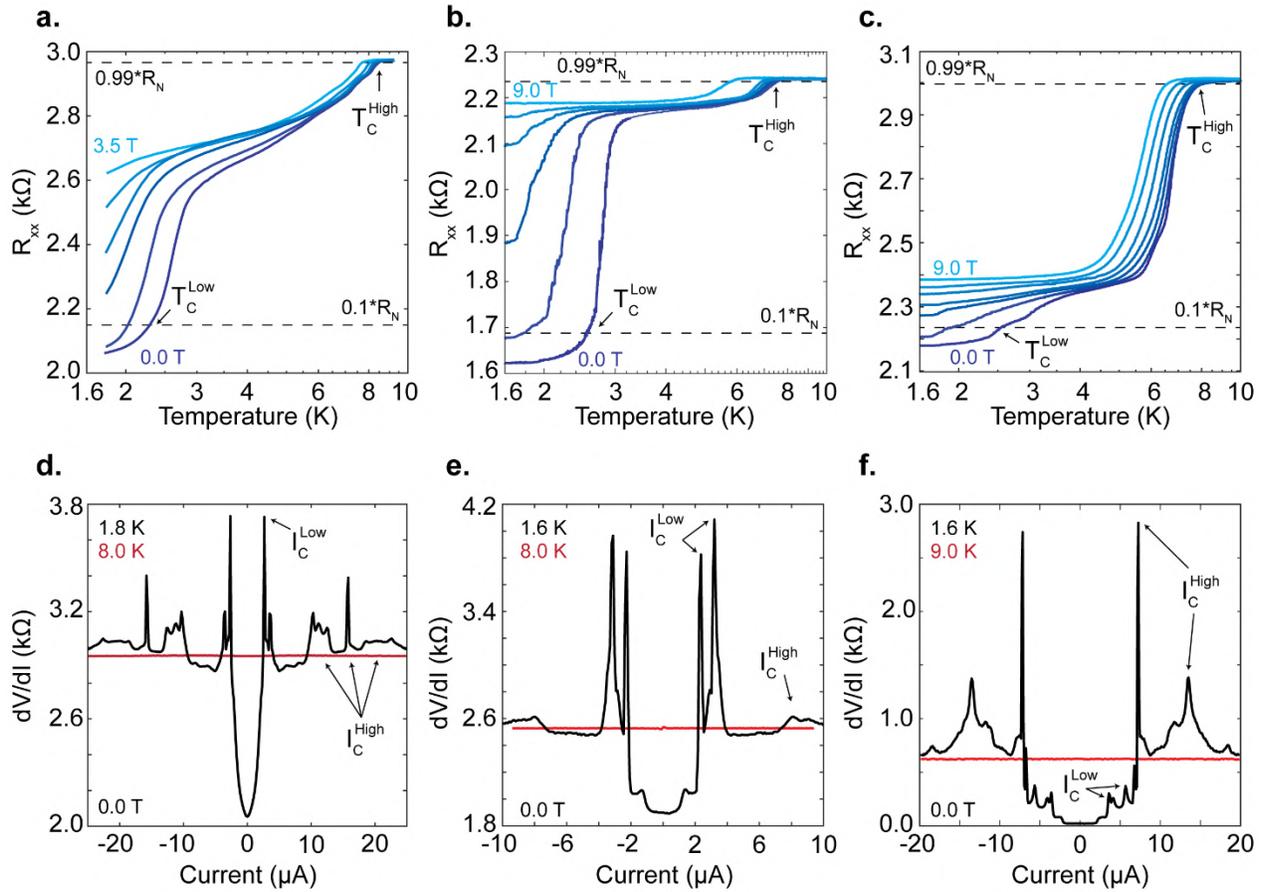

**Supplementary Figure 6. Superconductivity in various current annealed $Re_6Se_8Cl_2$ flakes. a, b, c,** Resistance vs. temperature at various magnetic fields for 3 pre-patterned $Re_6Se_8Cl_2$ devices. **d, e, f,** Differential resistance vs. DC bias in the superconducting state (black) and the normal state (red) for the same or similar devices as in (**a**), (**b**), and (**c**) respectively.

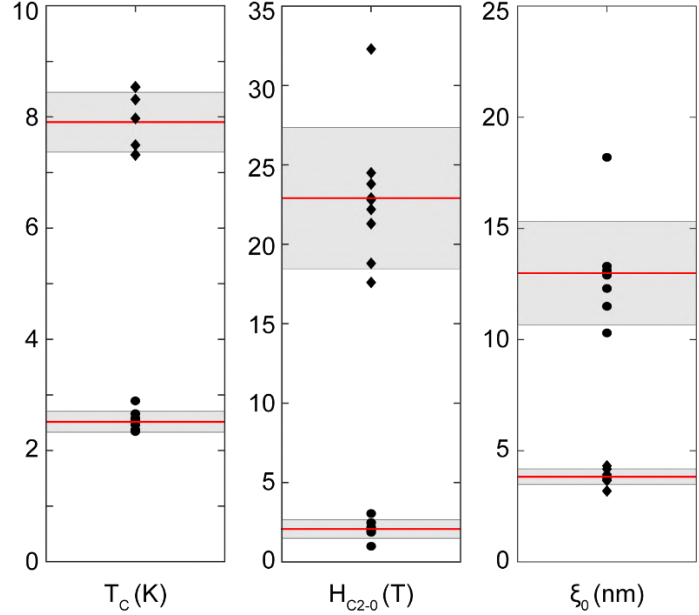

**Supplementary Figure 7. Statistics for all superconducting Re$_6$Se$_8$Cl$_2$ flakes.** $T_C^{Low}$, $H_{C2 0}^{Low}$, and $\xi_0^{Low}$ (filled black circles) and $T_C^{High}$, $H_{C2\text{-}0}^{High}$, and $\xi_0^{High}$ (filled black diamonds) are shown. Corresponding mean values and standard deviations are given by a solid red line and grey region respectively.

### 4.2 Angle dependence of the upper critical field

In Supplementary Figure 8, we examine the effect of magnetic field direction on the superconducting properties of Re$_6$Se$_8$Cl$_2$ flakes. Supplementary Figure 8a,d presents two devices which exhibit clear, sharp transitions for both $T_C^{Low}$ and $T_C^{High}$. We examine the magnetic field-temperature phase diagram for magnetic fields both perpendicular (Supplementary Figure 8b,e) and parallel (Supplementary Figure 8c,f) to the ab plane. We observed no significant difference between field directions, indicating the superconductivity in this system is not dominated by two-dimensional confinement effects[11].

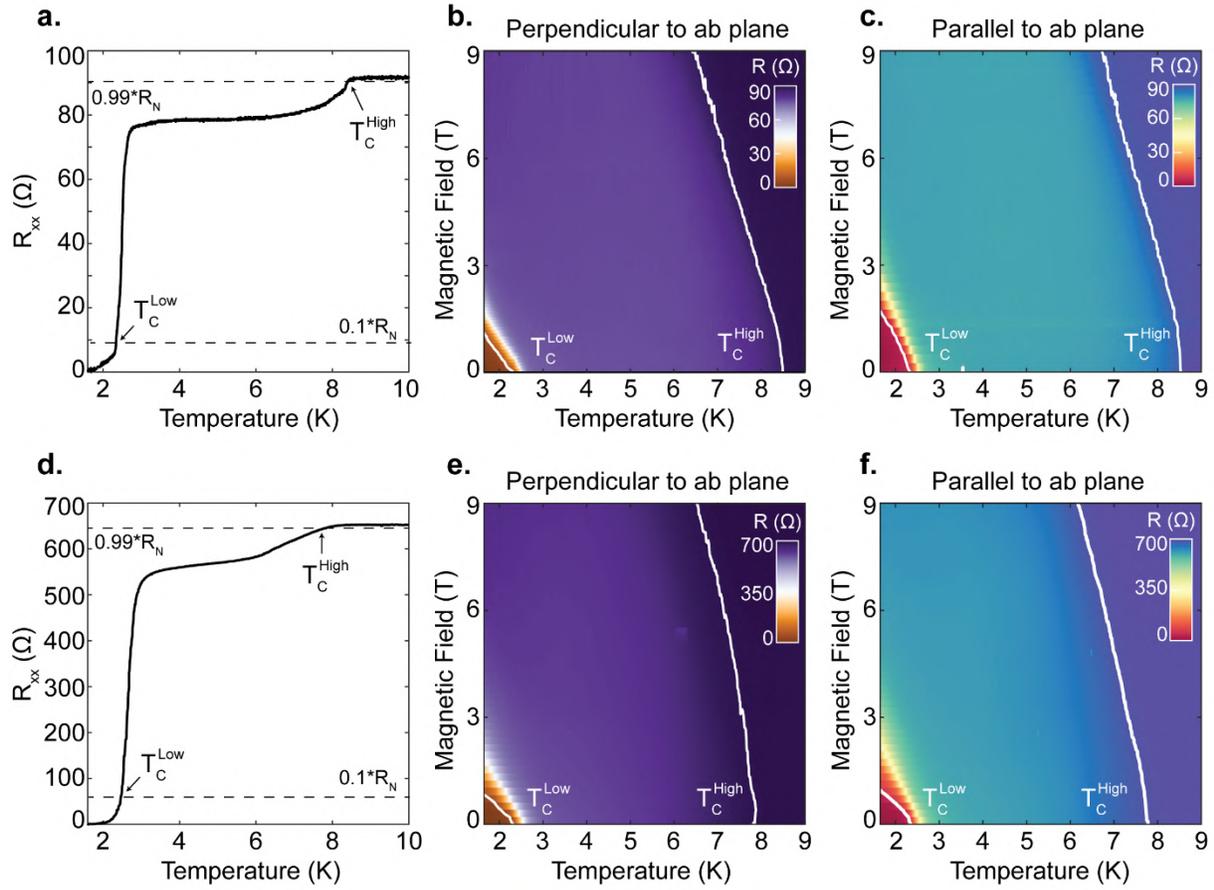

**Supplementary Figure 8. Angle dependence of the critical magnetic field in current annealed Re$_6$Se$_8$Cl$_2$. a,** Resistance vs. temperature at zero magnetic field for a current annealed Re$_6$Se$_8$Cl$_2$ flake. **b,** 2D color plot of resistance vs. temperature and magnetic field perpendicular to the ab plane for the flake in (**a**). **c,** 2D color plot of resistance vs. temperature and magnetic field parallel to the ab plane for the flake in (**a**). **d, e,** and **f,** Corresponding measurements in (**a**), (**b**), and (**c**) for a second Re$_6$Se$_8$Cl$_2$ flake. In all plots, $T_C^{Low}$ and $T_C^{High}$ are denoted. For all panels, a temperature independent background resistance was subtracted.

### 4.3 Vortex activation energy

Examining in detail the dependence of resistance vs. temperature and magnetic field for both devices presented in Supplementary Figure 8, we can fit both transitions to the Arrhenius equation, $R = R_0 e^{-\frac{U(B)}{k_B T}}$ (Supplementary Figure 9a,c for $T_C^{Low}$ and Supplementary Figure 9b,d for $T_C^{High}$). Note that in order to properly fit the Arrhenius equation, we need to subtract any offset resistance so the saturated regimes are set to zero resistance. The field dependent vortex activation, U(B) is well fit by a logarithmic field dependence, $U(B)=U_0\log(H_0/H)$, which is characteristic of layered type-II superconducting systems[12]. $U_0$ is the strength of vortex pinning and interaction and $H_0$ is the upper critical field, $H_{C2-0}$. Extracted $H_0$ values agree with our measured upper critical fields (Supplementary Figure 7).

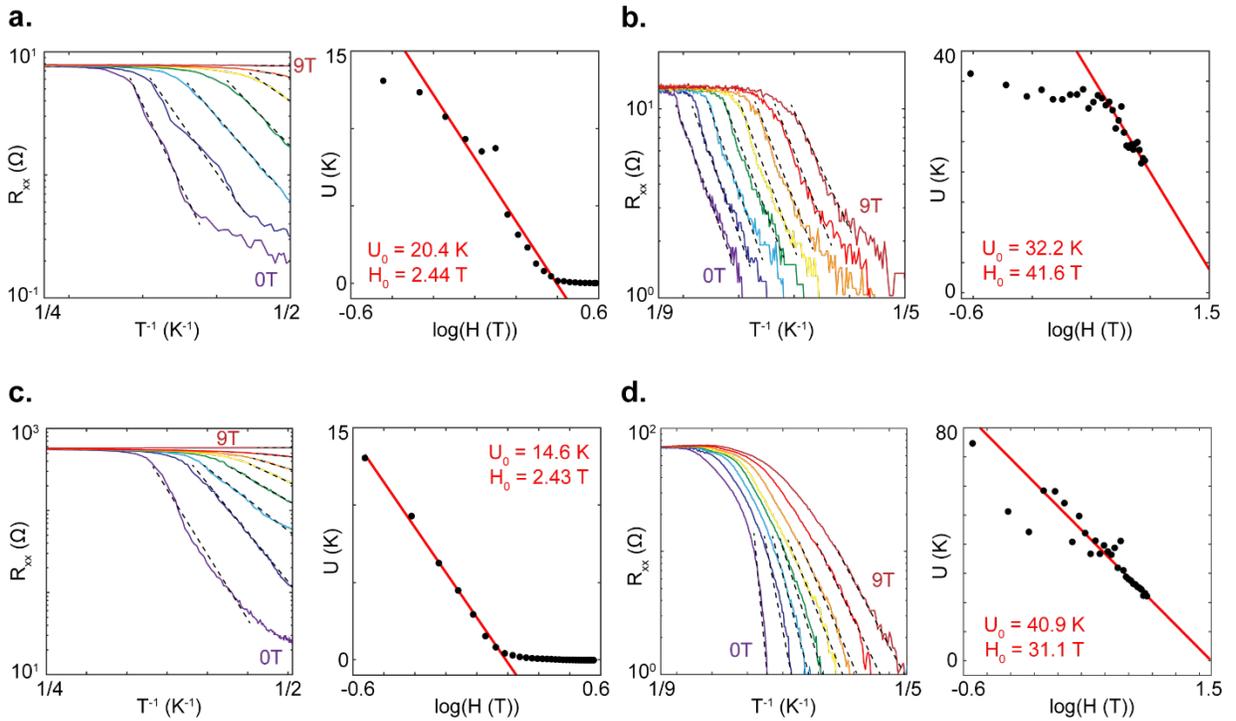

**Supplementary Figure 9. Thermal activation of vortices in current annealed $Re_6Se_8Cl_2$. a,b,** Resistance vs. inverse temperature on a log scale at various magnetic fields (left) and the extracted vortex activation energy vs. magnetic field on a log scale (right) for $T_C^{Low}$ (**a**) and $T_C^{High}$ (**b**) for the same device in Supplementary Figure 8a. **c,d,** Resistance vs. inverse temperature on a log scale at various magnetic fields (left) and the extracted vortex activation energy vs. magnetic field on a log scale (right) for $T_C^{Low}$ (**d**) and $T_C^{High}$ (**e**) for the same device in Supplementary Figure 8d.

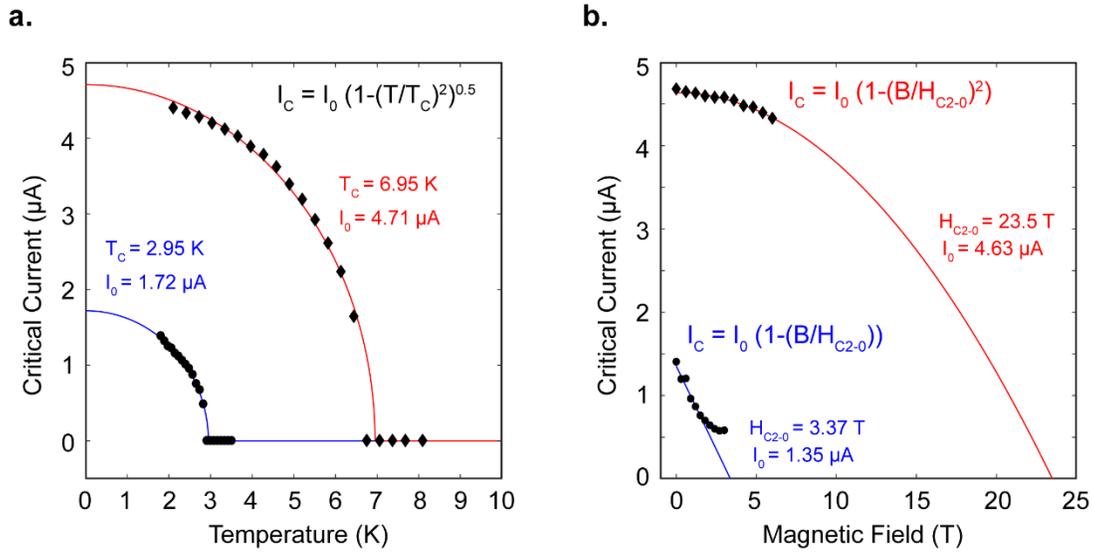

**Supplementary Figure 10. Fitting critical current vs. temperature and magnetic field. a,** Extracted $I_C^{High}$ (black diamonds) and $I_C^{Low}$ (black circles) vs. temperature for the device in Figure 2d. **b,** Extracted $I_C^{High}$ (black diamonds) and $I_C^{Low}$ (black circles) vs. magnetic field perpendicular to the cluster planes for the device in Figure 2d. The best fitting equations are shown with corresponding fit parameters[13]

### 4.4 Superconductivity in bulk current and thermally annealed Re$_6$Se$_8$Cl$_2$

We were able to achieve superconductivity through both current and thermal annealing in several bulk devices. Through thermal annealing (Supplementary Fig. 11c,d) we observed only a single transition in the resistance at low temperature. We believe this transition is due to metallic Re, which is created when annealing temperatures approach the synthesis temperature (>900 °C). Through current annealing we were able to achieve two transitions in certain devices, $T_C^{High}$ and $T_C^{Low}$ (Supplementary Fig. 11a), and only a single lower transition, $T_C^{Low}$, in others (Supplementary Fig. 11b).

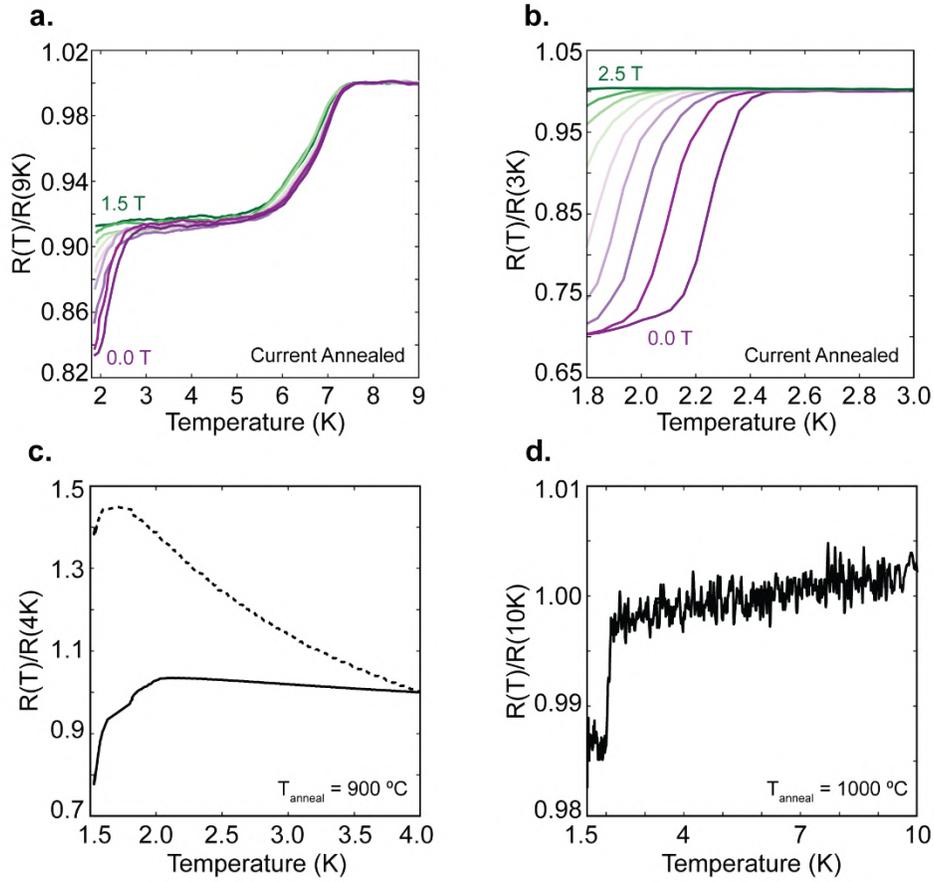

**Supplementary Figure 11. Superconducting properties of bulk single crystal annealed $Re_6Se_8Cl_2$. a,b,** Resistance vs. temperature at various magnetic fields for two current annealed bulk $Re_6Se_8Cl_2$ devices. Resistance is normalized to the 9 K (**a**) and 3 K (**b**) values. **c,d,** Resistance vs. temperature for bulk $Re_6Se_8Cl_2$ thermally annealed at 900 °C (**c**) and 1000 °C (**d**). Resistance is normalized to the 4 K (**c**) and 10 K (**d**) values.

## 5. Chemical and structural analysis

### 5.1 Details of thermal annealing

For all measurements which present single crystal and microcrystalline $Re_6Se_8Cl_2$ characterization vs. annealing temperature, we outline the details of the thermal annealing process. Individual (2-3) $Re_6Se_8Cl_2$ crystals or ~15 mg of microcrystalline powder is sealed in a fused silica tube under vacuum (~40 mTorr). The tube is placed in a box oven pre-heated to the selected temperature 200 - 1000 °C. The tube is removed after 1 hour and quenched in water.

### 5.2 FIB cut and TEM measurements

For transmission electron microscope (TEM) studies, focused ion-beam (FIB), FEI Helios NanoLab 660, was used to prepare a cross sectional foil. To protect the contacts against ion-beam damage during sample preparation, amorphous platinum (2 µm thick) was sputtered on the surface of the wafer by electron then ion beam. During initial steps of FIB sample preparation, an accelerating voltage of 30 kV was used to make a lamella of ~100 nm thickness. Further thinning was done at 5 kV for longer time with a gradually decreasing ion current to remove the damaged layers on each side and achieve an electron transparent foil. TEM and high-resolution scanning TEM (HR-STEM) analyses were performed on a FEI Talos F200X at an accelerating voltage of 200 kV. In order to achieve the highest resolution without damaging the sample, the spot size was set to 9 to reduce the electron beam current and the condenser 2 aperture was set to 50 µm to obtain a small beam-convergence angle.

**5.3 Scanning electron microscopy and energy dispersive x-ray spectroscopy**

Scanning electron micrographs of the devices were collected on a Zeiss Sigma VP SEM. EDX of the single crystal and microcrystalline powder samples was performed with a Bruker XFlash 6130 attachment. Spectra were collected with a beam energy of 15 keV. Elemental compositions and atomic percentages were estimated by integrating under the characteristic spectrum peaks for each element using Bruker ESPRIT 2 software. Atomic percentages were calculated assuming only Re, Se and Cl were present. To estimate the relative loss of Cl, the atomic percentage of Cl at each temperature was divided by the atomic percentage of Cl of the non-annealed sample. The same process was used for Se.

In Supplementary Figure 12, we examine Cl and Se concentration vs. annealing temperature for both single crystal and microcrystalline $Re_6Se_8Cl_2$ samples. We find a significant difference between single crystal and microcrystalline powder. Microcrystalline powder shows a decrease in Cl concentration at temperatures lower than that of the single crystal. We believe this is due to the dramatic difference in particle surface area and volume. In the main text (Fig. 4d), we showed that surface Cl desorbs at lower temperatures than bulk Cl. In this context, the high surface area microcrystalline powder will show a larger decrease in Cl concentration at the surface desorption temperature (~400 °C), whereas single crystals only show significant Cl decrease at the bulk desorption temperature (~800 °C).

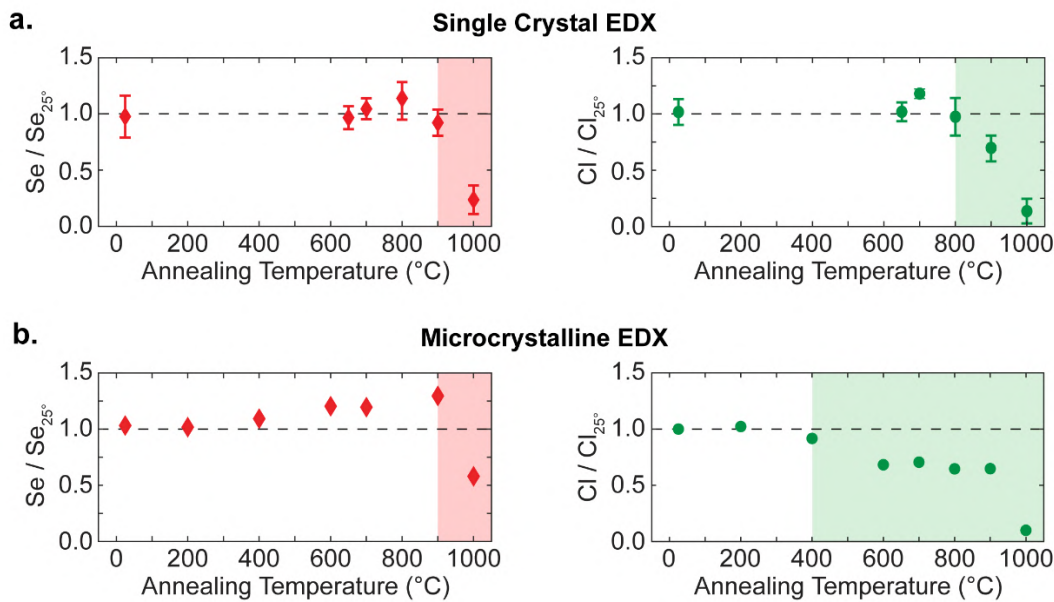

**Supplementary Figure 12. EDX vs. thermal annealing on microcrystalline and single crystal Re$_6$Se$_8$Cl$_2$. a,b,** Se and Cl composition measured in EDX vs. annealing temperature for Re$_6$Se$_8$Cl$_2$ single crystals (**a**) and microcrystalline Re$_6$Se$_8$Cl$_2$ powders (**b**).

### 5.4 Single crystal x-ray diffraction

We measured the Re$_6$Se$_8$Cl$_2$ single crystal lattice parameters after treatment at various thermal annealing temperatures. Single crystal x-ray diffraction data was collected on an Agilent SuperNova diffractometer using mirror-monochromated Mo Kα radiation. Each crystal was mounted under oil at room temperature using a MiTeGen MicroMount. Data collection and unit cell determination was performed in CrysAlisPro. We find no change in lattice parameters up to 1000 °C (Supplementary Figure 13).

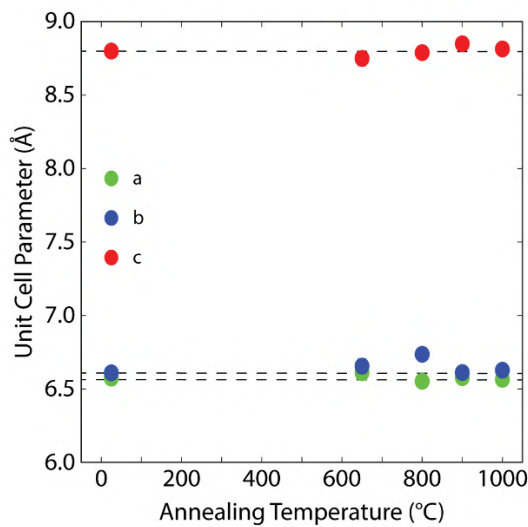

**Supplementary Figure 13. Single crystal x-ray diffraction vs. annealing temperature.** a, b, and c lattice parameters are shown vs. annealing temperature. Estimated uncertainties are within the area of the points.

### 5.5 X-ray photoemission spectroscopy

### 5.5.1 Microcrystalline powder

A small amount (~10 mg) of pristine or thermally annealed microcrystalline powder was pressed into In foil and mounted onto a chuck. The sample was measured using a Phi 5500 XPS with Mg source. Regions collected were Re 4f, Se 3d, and Cl 2p. Background subtraction, peak fitting and integration were performed using the XPST (X-ray Photoelectron Spectroscopy Tools) package (v.1.3) in Igor Pro. For each sample, the total area of each element's characteristic set of peaks was summed. To estimate the relative loss of Cl, the total Cl peak area was divided by the total Re peak area at each temperature, and this ratio was normalized to the untreated sample. The same process was used for Se. We present a summary of the XPS measurement results for select thermal annealing temperatures in Supplementary Figure 14.

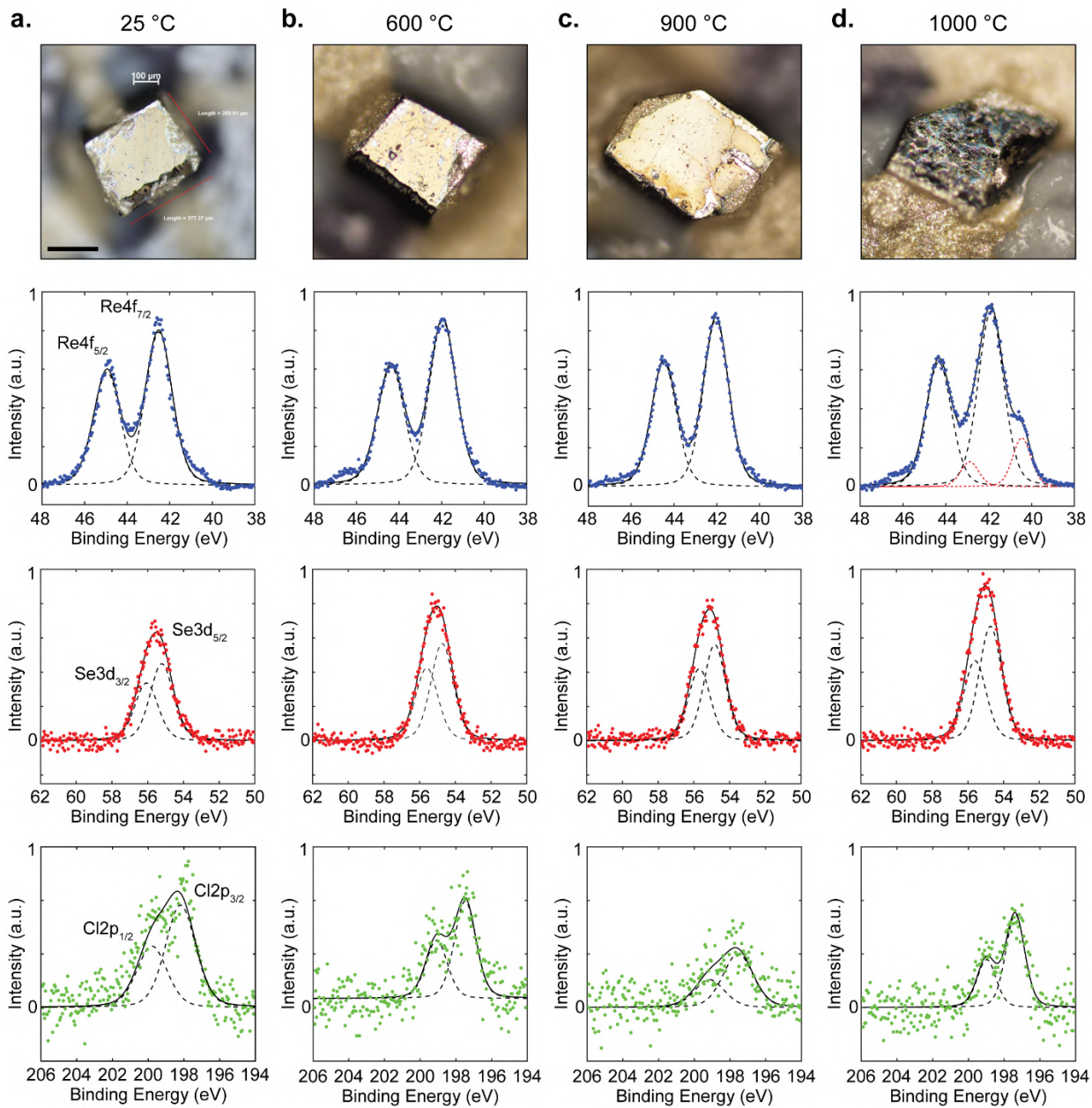

**Supplementary Figure 14. $Re_6Se_8Cl_2$ XPS spectra vs. thermal annealing temperature.** Optical image with corresponding XPS spectrum for Re, Se, and Cl for non-annealed (**a**), annealed at 600 °C (**b**), annealed at 900 °C (**c**), and annealed at 1000 °C (**d**). In all XPS spectra, dashed black lines represent single Gaussian fits and solid black lines represent the total spectra fits. Optical images are taken parallel to the c axis. The scale bar in (**a**) is 200 μm and applies to the optical images in (**b**),(**c**), and (**d**). In (**d**), the dashed red lines represent the Gaussian fits to Re metal signal.

### 5.5.2 Exfoliated flakes

In an attempt to better quantify the Cl loss after current annealing $Re_6Se_8Cl_2$ flakes, we performed μ-XPS on exfoliated $Re_6Se_8Cl_2$ flakes contacted on the top surface by metal electrodes embedded in hexagonal boron nitride (h-BN), called "window via contacts" (Supplementary Fig. 15a)[14]. This allows for the collection of μ-XPS spectra within a localized spatial region before (Supplementary Fig. 15c,e) and after current annealing the top surface (Supplementary Fig. 15d,f). μ-XPS measurements were performed using a Physical Electronics VersaProbe II XPS with Al source. The device was located using scatter x-ray imaging (SXI). An x-ray spot size of 15 μm was focused over the sample. Regions collected were Re 4f, Se 3d, and Cl 2p. Background subtraction, peak fitting and integration were performed using the XPST package (v.1.3) in Igor Pro. To estimate the relative loss of Cl after current annealing, the total Cl peak area was divided by the total Re peak area and normalized to the untreated sample (Supplementary Fig. 15h). Overall, we were able to measure a significant decrease in Cl concentration and sample resistance (Supplementary Fig. 15g) after current annealing. The same μ-XPS procedure was used to collect μ-XPS spectra on flipped current annealed $Re_6Se_8Cl_2$ flakes.

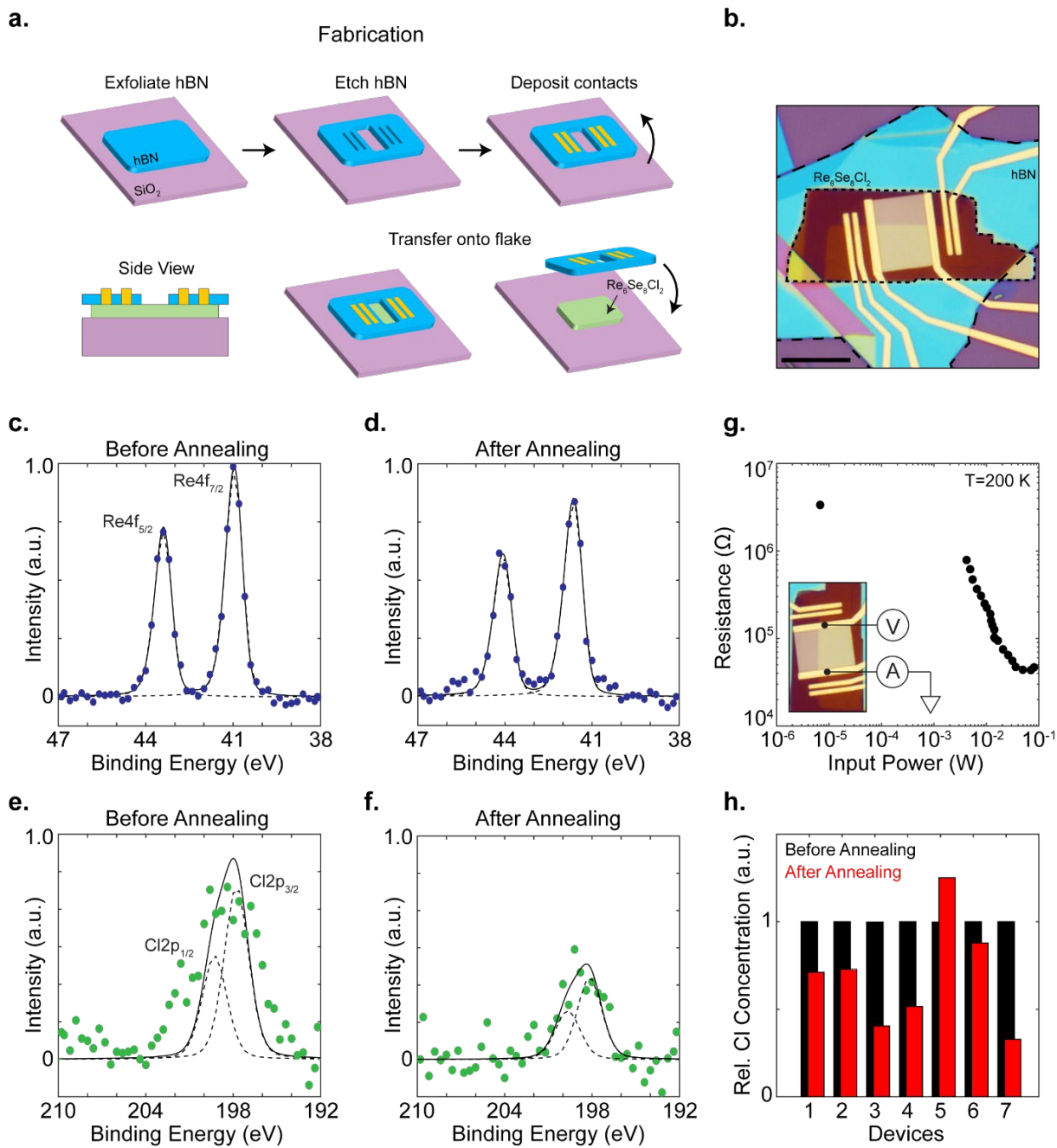

**Supplementary Figure 15: XPS on window via devices. a,** fabrication process of window via devices. **b,** Optical image of a typical window via device. Scale bar is 10 μm. **c,d** μ-XPS spectra of the Re binding energies before (**c**) and after (**d**) current annealing. **e,f** μ-XPS spectra of the Cl binding energies before (**e**) and after (**f**) current annealing. **g,** Resistance vs. current annealing power. **h,** Cl μ-XPS peak area normalized by Re μ-XPS peak area before (black) and after (red) annealing for 5 window via devices (1-5) and two reference top contacted flakes (6,7).

### 5.6 Raman spectroscopy

Raman spectra were taken in a Renishaw inVia using a 532 nm wavelength laser. A laser power of ~1 mW was used. Spectra were accumulated for 3 minutes with a grating of 2400 gr/mm. A 100x objective was used, with a spot size of ~1 - 2 μm.

### 5.7 Thermal annealing mass spectrometry

To perform a real time analysis of desorbing elements, we developed a technique dubbed thermal annealing mass spectrometry (TA-MS). A ceramic boat was filled with 100 mg of $Re_6Se_8Cl_2$ microcrystalline powder and placed in a long fused silica tube (50 cm x 2 cm ID). The tube was placed into a tube furnace with the sample located in the center with $N_2$ gas flowing through the tube at a pressure of 5 psi. The outlet of the tube was connected to an APCI source on an Advion CMS mass spectrometer (Supplementary Fig. 16). Volatiles were ionized using APCI (negative mode), and data were collected in selected ion monitoring (SIM) mode. For Cl monitoring, the m/z's for the ions $Cl^-$ and $Cl_2^-$ were recorded. For Se monitoring, the m/z's for the ions $Se^-$, $Se_2^-$, and $Se_3^-$ were recorded. The furnace was heated to 800 °C at a rate of 20 °C/min and then held at 800 °C for 30 minutes. For Se monitoring, the furnace was then further heated to 920 ºC at a rate of 7 ºC/min.

In Supplementary Figure 17, we show an extended plot of $Se^-$, $Se_2^-$, and $Se_3^-$ counts measured vs. temperature up to 920 ºC. We see the onset of Se desorption at ~850 °C, which is consistent with our EDX results.

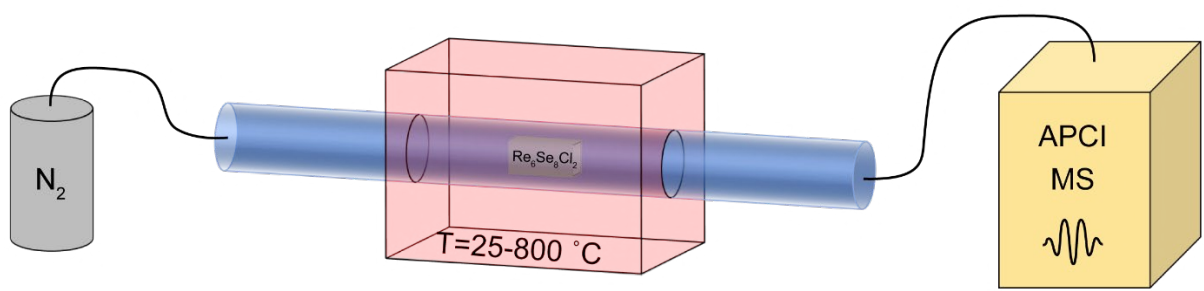

**Supplementary Figure 16. Thermal annealing mass spectrometry setup.** $N_2$ gas (gray) is flowed continuously through a quartz tube containing bulk $Re_6Se_8Cl_2$ crystals, gradually heated to a set temperature (blue, red). The output is flowed into a mass spectrometer (yellow).

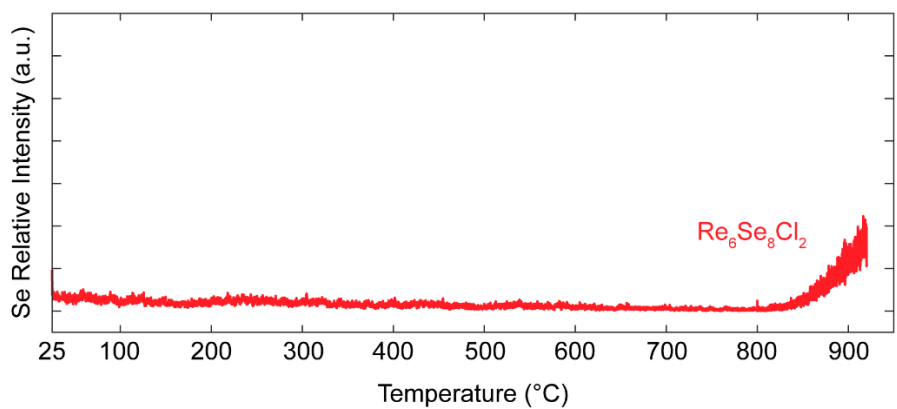

**Supplementary Figure 17. Thermal annealing mass spectrometry on $Re_6Se_8Cl_2$:** $Se^-$, $Se_2^-$, and $Se_3^-$ counts vs. sample temperature, measured in the TA-MS setup.

**References:**

1.  Novoselov, K. S. *et al.* Electric field effect in atomically thin carbon films. *Science* **306**, 666–9 (2004).

2.  Wang, L. *et al.* One-Dimensional Electrical Contact to a Two-Dimensional Material. *Science* **342**, 614–617 (2013).

3.  Asheghi, M., Touzelbaev, M. N., Goodson, K. E., Leung, Y. K. & Wong, S. S. Temperature-Dependent Thermal Conductivity of Single-Crystal Silicon Layers in SOI Substrates. *J. Heat Transfer* **120**, 30 (1998).

4.  Inyushkin, A. V., Taldenkov, A. N., Gibin, A. M., Gusev, A. V. & Pohl, H.-J. On the isotope effect in thermal conductivity of silicon. *Phys. status solidi* **1**, 2995–2998 (2004).

5.  Cahill, D. G., Watson, S. K. & Pohl, R. O. Lower limit to the thermal conductivity of disordered crystals. *Phys. Rev. B* **46**, 6131–6140 (1992).

6.  Powell, R. W., Tye, R. P. & Woodman, M. J. Thermal Conductivities and Electrical Resistivities of the Platinum Metals. *Platin. Met. Rev.* **6**, 138–143 (1962).

7.  Rosenberg, H. M. The Thermal Conductivity of Metals at Low Temperatures. *Philos. Trans. R. Soc. London. Ser. A, Math. Phys. Sci.* **247**, 441–497 (1955).

8.  Malen, J. A. *et al.* Optical Measurement of Thermal Conductivity Using Fiber Aligned Frequency Domain Thermoreflectance. *J. Heat Transfer* **133**, 081601 (2011).

9.  Ong, W.-L. *et al.* Orientational order controls crystalline and amorphous thermal

transport in superatomic crystals. *Nat. Mater.* **16**, 83–88 (2017).

10. Bergman, T. L. & Incropera, F. P. *Fundamentals of heat and mass transfer.* (Wiley, 2011).

11. Tinkham, M. *Introduction to superconductivity. Dover books on physics* (Dover Publ, 2004).

12. Feigel'man, M. V., Geshkenbein, V. B. & Larkin, A. I. Pinning and creep in layered superconductors. *Phys. C Supercond. its Appl.* **167**, 177–187 (1990).

13. Dew-Hughes, D. The critical current of superconductors: an historical review. *Low Temp. Phys.* **27**, 713–722 (2001).

14. Telford, E. J. *et al.* Via Method for Lithography Free Contact and Preservation of 2D Materials. *Nano Lett.* **18**, 1416–1420 (2018).